\documentclass[11pt]{article}
\pdfoutput=1
\usepackage{jheppub}
\usepackage{graphicx}
\usepackage{amssymb}
\usepackage{amsmath,amssymb}
\usepackage{slashed}
\usepackage{hyperref}
\usepackage{caption}
\usepackage{xcolor}
\usepackage{dsfont}
\usepackage{verbatim}
\usepackage{subfig}
\usepackage{amsmath,amssymb,amscd,amsfonts,mathtools}
\usepackage{xcolor}
\definecolor{markgreen}{RGB}{230,243,230}
\definecolor{darkolivegreen}{rgb}{0.33, 0.42, 0.18}
\definecolor{darkpastelgreen}{rgb}{0.01, 0.75, 0.24}

\DeclareMathOperator{\arcsinh}{arcsinh}

\usepackage[toc,page]{appendix}
\usepackage{epsfig}
\usepackage{epstopdf}
\usepackage{latexsym}
\usepackage{graphicx}
\usepackage{booktabs}
\usepackage{bbm}
\usepackage{color}
\usepackage{physics}
\usepackage{tensor}
\usepackage{verbatim}
\usepackage{subfig}
\usepackage{tikz}
\usepackage{ifthen}
\usetikzlibrary{matrix}
\usetikzlibrary{decorations.markings,calc,shapes,decorations.pathmorphing}
\usetikzlibrary{patterns}
\usetikzlibrary{positioning}

\pdfoutput=1
\makeatletter
\def\@fpheader{\relax}
\makeatother

\newboolean{showcomments}
\setboolean{showcomments}{false}
\newcommand\rem[1]{\ifthenelse{\boolean{showcomments}}{{#1}}{}}
\newcommand{\be}{\begin{equation}}
\newcommand{\ee}{\end{equation}}

\author{Hao Geng}
\affiliation{Harvard University, 17 Oxford St., Cambridge, MA, 02139, USA.}

\emailAdd{haogeng@fas.harvard.edu}
\title{\Large Aspects of AdS$_2$ Quantum Gravity and the Karch-Randall Braneworld}
\preprint{\today}
\abstract{
   In this paper, we use the Karch-Randall braneworld to study theories of quantum gravity in two dimensional (nearly) anti-de Sitter space (AdS$_2$). We focus on effective gravitational theories in the setup with two Karch-Randall branes embedded in an asymptotically AdS$_3$ bulk forming a wedge. We find the appearance of two-dimensional Einstein-Hilbert gravity (or the Lorenzian version of Marolf-Maxfield theory)
   when the branes are rigid but the emergence of a class of dilaton gravity models parameterized by the tensions of the two branes when brane fluctuations are accounted for. A special case of our result is Jackiw-Teitelboim (JT) gravity, which has been proven useful to address many important problems in quantum gravity. 
   An important implication of our work is that these models have holographic duals as one-dimensional quantum mechanics systems. At the end, we discuss a puzzle regarding the energy spectrum and its resolution.}

\begin{document}	
\maketitle
\flushbottom
\section{Introduction}
The Karch-Randall braneworld \cite{Karch:2000gx,Karch:2000ct} played an important role in  recent progress of our understanding about quantum gravity. This includes providing solvable toy models to calculate the \textit{Page curve} of black hole radiation in general dimensions \cite{Almheiri:2019yqk,Almheiri:2019psy,Geng:2020qvw,Geng:2021mic,Chen:2020uac,Chen:2020hmv}, revealing the fact that in higher dimensions (the spacetime dimension $D\geq3$) the \textit{entanglement island}  exists  only in massive gravity theories for a very generic class of spacetimes \cite{Geng:2020qvw,Geng:2020fxl,Geng:2021hlu}, providing holographic models to study fine-grained properties of quantum field theory in black hole background \cite{Geng:2021hlu,Geng:2022yyy} and motivating a new holographic setup to study quantum gravity in anti-de Sitter space (AdS)-- the \textit{wedge holography} \cite{Akal:2020wfl,Miao:2020oey,Geng:2020fxl,Geng:2021iyq}. 

The original Karch-Randall braneworld \cite{Karch:2000ct,Karch:2000gx} describes the gravitational theory in AdS$_{d+1}$ (more precisely asymptotically AdS$_{d+1}$) with a codimension-one brane embedded. The brane geometry is (asymptotically) AdS$_d$ and the bulk gravitational theory is  Einstein-Hilbert gravity in the semiclassical limit. The localized gravity theory on the brane is quantum. However, in this setup it is hard to study 2d quantum gravity by embedding a two dimensional Karch-Randall brane into an AdS$_3$ bulk.  This is because there expects to be no localized graviton on the brane \cite{Karch:2000ct}.\footnote{The transverse traceless gauge removes all components of the 2d graviton.}

In this paper, we will show that the wedge holography setup provides a portal through which we can study two-dimensional quantum gravity by realizing it as the effective description for the intermediate picture of the setup. 
In this setup, we simply embed two AdS$_d$ branes into the AdS$_{d+1}$ bulk (see \cite{,Miao:2021ual} for a further generalization). 
The two branes intersect each other only on the asymptotic boundary and so they form a wedge in the AdS$_{d+1}$ bulk. This setup is interesting as it is a co-dimension two holographic setup and it has the following three equivalent descriptions \cite{Geng:2020fxl}:
\begin{enumerate}
\item \textbf{Bulk description}: Einstein gravity in an AdS$_{d+1}$ space $\mathcal{M}'_{d+1}$ containing two AdS$_d$ branes $\mathcal{M}_d^{(L)}$ and $\mathcal{M}_d^{(R)}$,which intersect each other on the asymptotic boundary (we call the place they intersect as the defect and it is (d-1)-dimensional) and therefore form a \textit{wedge};
\item \textbf{Intermediate description}: Two $d$-dimensional CFTs coupled to gravity on distinct asymptotically AdS$_d$ spaces $\mathcal{M}_d^{L}$ and $\mathcal{M}_d^{R}$, with these systems being connected via a transparent boundary condition at a defect $\mathcal{M}_{d-1}^{(0)}=\partial\mathcal{M}_d^{L}=\partial\mathcal{M}_{d}^{R}$;
\item \textbf{Boundary description}: A $(d-1)$-dimensional CFT on $\mathcal{M}_{d-1}^{(0)}$. 
\end{enumerate}
We exclusively consider the case $d=2$ in this paper. Even though there are no 
brane-localized graviton modes in this new setup,  the two branes embedded in such a special way that the bulk is compact in the direction perpendicular to the brane so there exists a lower-dimensional graviton that is confined in the wedge. The aim of this paper is to study the induced two dimensional quantum gravity based on this observation. 

We start with the analysis of the holographic entanglement entropy in  AdS$_3$ wedge holography. Unlike the higher dimensional cases studied in \cite{Geng:2020fxl}, the defect of in the AdS$_3$ case has two disconnected components even when the bulk geometry is empty AdS$_3$ (in global coordinates). This is a purely topological effect. We study both the entanglement entropy between these two disconnected components and the L/R entropy we introduced in \cite{Geng:2020fxl}, which probes the dynamics of the holographic dual system. For the first quantity we find that it is the same when the defect system is in the vacuum state and in the thermofield double state. Moreover, we find that there is an infinite degeneracy in the bulk for \textit{Ryu-Takayanagi} (RT) surface \cite{Ryu:2006bv} which calculates the entanglement entropy between the two defects. Hence this tells us that there is no definite RT surface and if this were the full answer, this would give indefinite entanglement wedges for the defects. We suggest that the correct interpretation of this apparent puzzle is that the system has trivial dynamics due to conformal symmetry. We provide further evidence for this suggestion and later precisely prove the suggestion by calculating the energy spectrum.

We then construct 2d quantum gravity theories with nontrivial dynamics in the context of  AdS$_3$ wedge holography. We carefully take into account the brane fluctuations which are ignored in the original wedge holography setup. We focus on the low energy dynamics of the brane fluctuation coupled to the 2d gravity confined in the wedge. We find that this 2d low energy effective action is precisely the Jackiw-Teitelboim (JT) gravity \cite{Teitelboim:1983ux,Jackiw:1984je,Fukuyama:1985gg} minimally coupled to a massive scalar field with the dilaton playing the role of the radion that describes the relative fluctuation of the two branes.

We also provide two special situations where we have pure JT gravity. As  was pointed out in \cite{Jensen:2016pah,Maldacena:2016upp,Engelsoy:2016xyb}, to see  nontrivial dynamics of JT gravity it is necessary to generate a nontrivial dilaton profile by introducing a UV cutoff\footnote{More precisely, we introduce a small UV cutoff $\epsilon$ and send it to zero at the end. This is what we mean by a UV cutoff in this paper. For studies of JT gravity with a finite cutoff we refer to \cite{Iliesiu:2020zld,Stanford:2020qhm,Griguolo:2021wgy}.} of the 2d spacetime which breaks the conformal symmetry and the emergent dynamics is that of the associated soft (Goldstone) modes.\footnote{It deserves to be mentioned that in the usual applications of the AdS/CFT correspondence we always have to introduce a small UV cutoff to the AdS geometry as many quantities like the entanglement entropy are not well-defined in CFT without such a cutoff. However, here in JT gravity the appearance of the UV cutoff is subtler and more interesting. This is due to the appearance of the dilaton field whose boundary profile introduces an explicit breaking of the asymptotic conformal symmetry and induces the nontrivial dynamics of the soft modes (see \cite{Maldacena:2016upp} or the later discussions of our current paper). We thank Hong Liu for pointing out this interesting aspect.} We will show that with the conformal symmetry breaking the degeneracy of the RT surfaces is lifted. Interestingly, We will show that our framework implies that this class of dilaton gravity theories have the holographic dual as random matrix theories if we assume gauge/gravity duality in the 3d bulk.

This paper is organized as follows. In Sec.~\ref{sec:review} we review relevant previous work \cite{Geng:2020fxl,Akal:2020wfl} on entanglement entropy in wedge holography. In Sec.~\ref{sec:EE} we study holographic entanglement entropies in 3d wedge holography by which we find an infinite degeneracy of RT surfaces. This causes an apparent puzzle regarding the entanglement wedge reconstruction in wedge holography. We comment on the correct interpretation and resolution of this apparent puzzle. In Sec.~\ref{sec:spectrumEE} we show that we can deduce the exact energy spectrum of the holographic dual quantum mechanics system from the study of entanglement entropy in Sec.~\ref{sec:EE}. In Sec.~\ref{sec:effactionJT} we include the brane fluctuation that is ignored in the original wedge holography setup \cite{Akal:2020wfl,Geng:2020fxl} in 3d and show that the low energy effective description in the intermediate picture is a class of dilaton gravity theories with JT gravity as a special case. In Sec.~\ref{sec:confbreaking} we show that the apparent puzzle caused by the infinite degeneracy of RT surfaces is lifted when the conformal symmetry is broken. In Sec.~\ref{sec:holographicdual} we speculate that our setup implies that this class of dilaton gravity theories have holographic dual as random matrix theories. In Sec.~\ref{sec:spectrumpuzzle} we point out a puzzle regarding the energy spectrum of the dilaton gravity, we show that the energy spectrum of the dilaton gravity will not reduce to that of the original 3d wedge holography in appropriate limit and we provide a resolution of this puzzle. In Sec.~\ref{sec:conclusion} we conclude our paper with discussions. In Appendix.~\ref{sec:bdyterms} we show the derivation of the boundary terms for the 2d dilaton gravity from the 3d Gibbons-Hawking term.

\section{Review of Previous Work}\label{sec:review}

In this section we review the work \cite{Geng:2020fxl,Akal:2020wfl} relevant to our current study. 
\subsection{Wedge Holography}
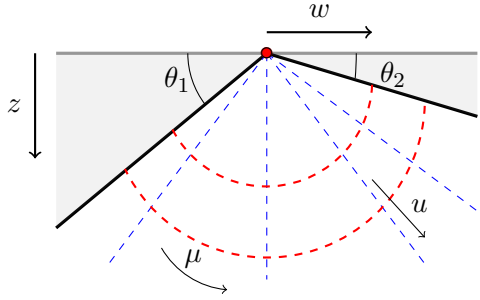
\begin{figure}[h]
\begin{centering}
\begin{tikzpicture}[scale=1.4]
\draw[-,very thick,black!40] (-2,0) to (0,0);
\draw[-,very thick,black!40] (0,0) to (2,0);

\draw[-,very thick] (0,0) to (2,-0.6);

\draw[-,very thick] (0,0) to (-2,-1.66);

\draw[fill=gray, draw=none, fill opacity = 0.1] (0,0) to (-2,-1.66) to (-2,0) to (0,0);
\draw[fill=gray, draw=none, fill opacity = 0.1] (0,0) to (2,0) to (2,-0.6) to (0,0);

\draw[-,dashed,color=blue] (0,0) to (-1.5,-2); 
\draw[-,dashed,color=blue] (0,0) to (0,-2.15);
\draw[-,dashed,color=blue] (0,0) to (1.5,-2); 
\draw[-,dashed,color=blue] (0,0) to (2,-1.5); 

\draw[-] (-0.75,0) arc (180:217.5:0.8);
\node at (-0.85,-0.25) {$\theta_1$};

\draw[-] (.85,0) arc (0:-4.95:2.75);
\node at (1.2,-0.2) {$\theta_2$};

\draw[-,dashed,color=red, thick] (1,-0.3) arc (-2.25:-90-52.5-10:1.005);
\draw[-,dashed,color=red, thick] (1.5,-0.5) arc (-2.25:-90-52.5-11.5:1.5);
\node at (0,0) {\textcolor{red}{$\bullet$}};
\node at (0,0) {\textcolor{black}{$\circ$}};

\draw[->,thick,color=black] (0.0,0.2) to (1,0.2);
\node at (0.5,0.4)
{\textcolor{black}{$w$}};

\draw[->,thick,color=black] (-2.2,0) to (-2.2,-1);
\node at (-2.4,-0.5)
{\textcolor{black}{$z$}};



\draw[->] (-1.0,-1.85) arc (210:265:0.8);
\node at (-0.7,-1.95) {$\mu$};

\draw[->] (1.0,-1.2) to (1.5, -1.75);
\node at (1.45,-1.45) {$u$};

\end{tikzpicture}
\caption{\small{\textit{An illustration of wedge holography: the bulk geometry is a part of empty AdS$_{d+1}$ between two end-of-world Karch-Randall branes (black lines), meeting each other on the conformal boundary of AdS$_{d+1}$ at their common boundary (the red dot). The bulk shaded regions behind the branes are removed, leaving only a wedge in the bulk. Wedge holography states that the gravitational physics in the bulk wedge is dual to a conformal field theory living on the red dot (the defect).}}}
\label{pic:wedgeholography}
\end{centering}
\end{figure}

The basic setup is  wedge holography, which is illustrated in Fig.\ref{pic:wedgeholography}. For simplicity, we start with an empty AdS$_5$ geometry in the Poincar\'{e} patch:
\begin{equation}
ds^2 = \frac{1}{z^2}(-dt^2 + dz^2 + d\vec{x}^2 + dw^2)\,,\label{ppatch}
\end{equation}
where we set the AdS curvature length to one, $w\in\mathbb{R}$ denotes the horizontal direction in Fig.~\ref{pic:wedgeholography}, and z is the radial coordinate such that the asymptotic boundary is $z\rightarrow0$. Then we embed two Karch-Randall branes into the bulk with positive tension and the two branes intersect each other at a defect on the asymptotic boundary of the bulk AdS$_{5}$. The defect is 3-dimensional. To easily describe the geometry of the branes we do the following coordinate transformation
\begin{equation}
z = u\sin\mu\,, \quad w = -u\cos\mu\,\label{coc},
\end{equation}
which transforms the bulk metric Equ.~(\ref{ppatch}) to
\begin{equation}
    ds^2=\frac{1}{\sin^{2}\mu}\left(\frac{-dt^2+d\vec{x}^2+du^2}{u^2}+d\mu^2\right)\,,\label{metric1}
\end{equation}
and we have $0 < \mu < \pi\,,u>0$.
The two branes are situated as the $\mu=\theta_{1}$ and the $\mu=\pi-\theta_2$ slices, their geometries are just empty AdS$_4$ and they cut off the bulk region behind them (the grey regions in Fig.~\ref{pic:wedgeholography}). 

Using  AdS/CFT holography for Karch-Randall braneworlds \cite{Maldacena:1997re,Karch:2000ct,Duff:2004wh}, it is easy to see that this configuration is doubly-holographic. This is because  AdS$_{5}$ quantum gravity can be firstly dualized to UV-cutoff conformal field theory on AdS$_4$ coupled with gravity.\footnote{More precisely, the gravity dynamics on AdS$_4$ is induced by the UV-cutoff CFT. This is a universal fact for braneworld gravity without the so-called DGP term. See for example \cite{Compere:2008us} for a discussion of how the gravity is induced in some detail.} Then as a quantum gravitational theory in the AdS$_{4}$ it is further dualized to a conformal field theory living on its asymptotic boundary which is
 the defect in Fig.~\ref{pic:wedgeholography}. Hence, we get a duality between a quantum gravitational theory in the AdS$_{5}$ wedge to a 3-dimensional conformal field theory on the boundary defect by applying holography twice. This argument persists for any asymptotically AdS$_{d+1}$ bulk geometry and there the geometry of the Karch-Randall brane is determined by the Israel's junction equation \cite{Israel:1967zz,Kraus:1999it} 
\begin{equation}
    K_{ab}=(K-T)h_{ab}\,,\label{eq:junction1}
\end{equation}
where $h_{ab}$ is the induced metric on the brane, $K_{ab}$ is the extrinsic curvature of the brane with $K$ as its trace under $h_{ab}$ and $T$ is the brane tension.

\subsection{The Entanglement Entropy of the Defect System}
In \cite{Geng:2020fxl} we computed the entanglement entropy of the defect system when the bulk is Einstein's gravity in empty AdS$_5$ as well as in the AdS$_5$ black string. We used the island rule in double-holography which is equivalent to the Ryu-Takayanagi proposal in wedge holography proposed in \cite{Akal:2020wfl}. 

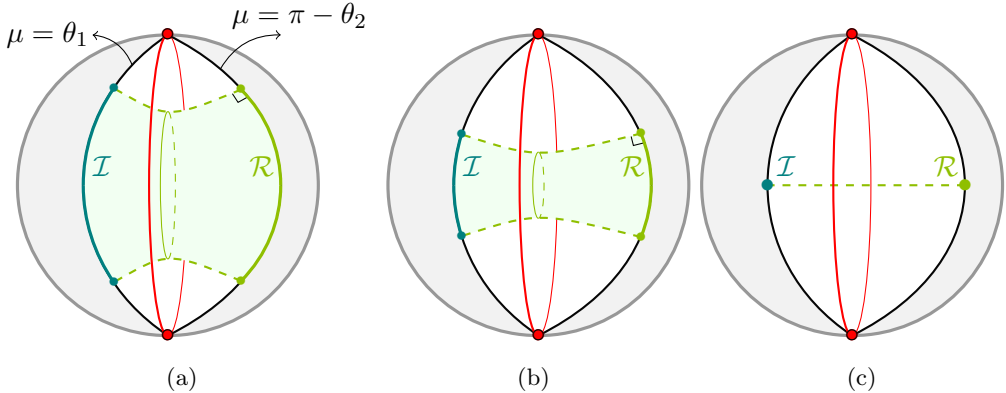
\begin{figure}
\centering
\subfloat[]{
\begin{tikzpicture}
\draw[-,red] (0,2) arc (90:-90:0.25 and 2);

\draw[-,very thick,black!40] (0,0) circle (2);
\draw[-,thick] (0,2) .. controls (-1.5,1) and (-1.5,-1) .. (0,-2);
\draw[-,thick] (0,2) .. controls (2,1) and (2,-1) .. (0,-2);

\draw[fill=gray, draw=none, fill opacity = 0.1] (0,-2) arc (-90:-270:2) .. controls (-1.5,1) and (-1.5,-1) .. (0,-2);

\draw[fill=gray, draw=none, fill opacity = 0.1] (0,-2) arc (270:450:2) .. controls (2,1) and (2,-1) .. (0,-2);

\draw[draw=none,fill=green!5] (1.5-1.8+1.2728,1.2728) .. controls (0,0.875) .. (-1.125+2.25-1.8431,1.2905) arc (145:215:2.25) .. controls (0,-0.875) .. (1.5-1.8+1.2728,-1.2728) arc (-45:45:1.8);

\draw[-] (1.5-1.8+1.2728+0.095,1.2728-0.095) to (1.5-1.8+1.2728-0.04,1.2728-0.17) to (1.5-1.8+1.2728-0.115,1.2728-0.055);

\draw[-,thick,color=black!25!lime,dashed] (1.5-1.8+1.2728,1.2728) .. controls (0,0.875) .. (-1.125+2.25-1.8431,1.2905);
\draw[-,thick,color=black!25!lime,dashed] (1.5-1.8+1.2728,-1.2728) .. controls (0,-0.875) .. (-1.125+2.25-1.8431,-1.2905);

\draw[black!25!lime] (0,0.98) arc (90:270:0.1 and 0.98);
\draw[black!25!lime,dashed] (0,0.98) arc (90:-90:0.1 and 0.98);

\draw[-,very thick,black!25!lime] (1.5,0) arc (0:45:1.8);
\draw[-,very thick,black!25!lime] (1.5,0) arc (0:-45:1.8);
\node at (1.25,0.25) {\textcolor{black!25!lime}{$\mathcal{R}$}};
\node[black!25!lime,scale=0.75] at (1.5-1.8+1.2728,1.2728) {$\bullet$};
\node[black!25!lime,scale=0.75] at (1.5-1.8+1.2728,-1.2728) {$\bullet$};

\draw[-,very thick,teal] (-1.125,0) arc (180:145:2.25);
\draw[-,very thick,teal] (-1.125,0) arc (180:215:2.25);
\node at (-0.875,0.25) {\textcolor{teal}{$\mathcal{I}$}};
\node[teal,scale=0.75] at (-1.125+2.25-1.8431,1.2905) {$\bullet$};
\node[teal,scale=0.75] at (-1.125+2.25-1.8431,-1.2905) {$\bullet$};

\draw[-,red,thick] (0,2) arc (90:270:0.25 and 2);
\node at (0,2) {\textcolor{red}{$\bullet$}};
\node at (0,2) {\textcolor{black}{$\circ$}};
\node at (0,-2) {\textcolor{red}{$\bullet$}};
\node at (0,-2) {\textcolor{black}{$\circ$}};
\draw[->] (-0.475,1.6) to[bend right] (-1,2);
\node at (-1.6,2) {$\mu = \theta_1$};
\draw[->] (0.7,1.55) to[bend left] (1.5,2);
\node at (1.75,2.25) {$\mu = \pi-\theta_2$};
\end{tikzpicture}}
\subfloat[]{
\begin{tikzpicture}
\draw[-,very thick,black!40] (0,0) circle (2);
\draw[-,thick] (0,2) .. controls (-1.5,1) and (-1.5,-1) .. (0,-2);
\draw[-,thick] (0,2) .. controls (2,1) and (2,-1) .. (0,-2);

\draw[-,red] (0,2) arc (90:-90:0.25 and 2);

\draw[fill=gray, draw=none, fill opacity = 0.1] (0,-2) arc (-90:-270:2) .. controls (-1.5,1) and (-1.5,-1) .. (0,-2);

\draw[fill=gray, draw=none, fill opacity = 0.1] (0,-2) arc (270:450:2) .. controls (2,1) and (2,-1) .. (0,-2);

\draw[draw=none,fill=green!5] (-0.3+1.6630,0.6888) .. controls (0,0.35) .. (1.125-2.1459,0.6766) arc (162.5:197.5:2.25) .. controls (0,-0.35) .. (-0.3+1.6630,-0.6888) arc (-22.5:22.5:1.8);

\draw[-] (-0.3+1.6630+0.06,0.6888-0.12) to (-0.3+1.6630-0.1,0.6888-0.16) to (-0.3+1.6630-0.13,0.6888-0.04);

\draw[-,thick,color=black!25!lime,dashed] (-0.3+1.6630,0.6888) .. controls (0,0.35) .. (1.125-2.1459,0.6766);
\draw[-,thick,color=black!25!lime,dashed] (-0.3+1.6630,-0.6888) .. controls (0,-0.35) .. (1.125-2.1459,-0.6766);

\draw[black!25!lime] (0,0.44) arc (90:270:0.075 and 0.44);
\draw[black!25!lime,dashed] (0,0.44) arc (90:-90:0.075 and 0.44);

\draw[-,very thick,black!25!lime] (1.5,0) arc (0:22.5:1.8);
\draw[-,very thick,black!25!lime] (1.5,0) arc (0:-22.5:1.8);
\node at (1.25,0.25) {\textcolor{black!25!lime}{$\mathcal{R}$}};
\node[black!25!lime,scale=0.75] at (-0.3+1.6630,0.6888) {$\bullet$};
\node[black!25!lime,scale=0.75] at (-0.3+1.6630,-0.6888) {$\bullet$};

\draw[-,very thick,teal] (-1.125,0) arc (180:162.5:2.25);
\draw[-,very thick,teal] (-1.125,0) arc (180:197.5:2.25);
\node at (-0.875,0.25) {\textcolor{teal}{$\mathcal{I}$}};
\node[teal,scale=0.75] at (1.125-2.1459,0.6766) {$\bullet$};
\node[teal,scale=0.75] at (1.125-2.1459,-0.6766) {$\bullet$};

\draw[-,red,thick] (0,2) arc (90:270:0.25 and 2);
\node at (0,2) {\textcolor{red}{$\bullet$}};
\node at (0,2) {\textcolor{black}{$\circ$}};
\node at (0,-2) {\textcolor{red}{$\bullet$}};
\node at (0,-2) {\textcolor{black}{$\circ$}};
\end{tikzpicture}}
\subfloat[]{
\begin{tikzpicture}
\draw[-,very thick,black!40] (0,0) circle (2);
\draw[-,thick] (0,2) .. controls (-1.5,1) and (-1.5,-1) .. (0,-2);
\draw[-,thick] (0,2) .. controls (2,1) and (2,-1) .. (0,-2);

\draw[-,red] (0,2) arc (90:-90:0.25 and 2);

\draw[fill=gray, draw=none, fill opacity = 0.1] (0,-2) arc (-90:-270:2) .. controls (-1.5,1) and (-1.5,-1) .. (0,-2);

\draw[fill=gray, draw=none, fill opacity = 0.1] (0,-2) arc (270:450:2) .. controls (2,1) and (2,-1) .. (0,-2);

\draw[-,thick,dashed,color=black!25!lime] (1.5,0) to (-1.125,0);

\node at (1.5,0) {\textcolor{black!25!lime}{$\bullet$}};
\node at (1.25,0.25) {\textcolor{black!25!lime}{$\mathcal{R}$}};
\node at (-1.125,0) {\textcolor{teal}{$\bullet$}};
\node at (-0.875,0.25) {\textcolor{teal}{$\mathcal{I}$}};

\draw[-,red,thick] (0,2) arc (90:270:0.25 and 2);
\node at (0,2) {\textcolor{red}{$\bullet$}};
\node at (0,2) {\textcolor{black}{$\circ$}};
\node at (0,-2) {\textcolor{red}{$\bullet$}};
\node at (0,-2) {\textcolor{black}{$\circ$}};

\node[color=white] at (1.6,2) {$\mu = \theta_1$};
\end{tikzpicture}
}
\caption{A constant-$t$ slice of AdS$_{d+1}$ foliated by global AdS$_d$ slices with two branes present. The defect is shown in red, and various candidate extremal surfaces (in order of decreasing area from left to right) are in green. For each surface, $\mathcal{R}$ and $\mathcal{I}$ are respectively the radiation region and island, which end orthogonally on both branes. The minimal, zero-area entanglement surface $r(\mu) = 0$ is shown on the right as a limit, cutting through the middle of the space.}
\label{pic:emptyads5}
\end{figure}

In the former case, it is easier to work in the global patch of those AdS$_4$ slices and the bulk geometry is shown in Fig.~\ref{pic:emptyads5}. The task translates to looking for a codimension-two entangling surface in the 5-dimensional bulk which is a minimal area surface, ends on the branes perpendicularly and is homologous to the boundary defect (a putative example of such a surface is shown in Fig.~\ref{pic:emptyads5}(a)). We find that in this case the entangling surface is given in Fig.~\ref{pic:emptyads5} (c) which degenerates and has zero area. Hence when the bulk is empty AdS$_5$ the wedge holographic dual has zero entanglement entropy which says that it is in a pure state consistent with the standard AdS/CFT correspondence. 

\begin{figure}
\begin{centering}
\begin{tikzpicture}[scale=1.4]
\draw[-,very thick,black!40] (-2,0) to (0,0);
\draw[-,very thick,black!40] (0,0) to (2,0);

\draw[-,very thick] (0,0) to (1,-0.1);

\draw[-,very thick] (0,0) to (-0.81,-0.62);

\draw[fill=gray, draw=none, fill opacity = 0.1] (0,0) to (-2,-1.5) to (-2,0) to (0,0);
\draw[fill=gray, draw=none, fill opacity = 0.1] (0,0) to (2,0) to (2,-0.2) to (0,0);

\draw[-,dashed,color=black!50] (0,0) to (-1.5,-2); 
\draw[-,dashed,color=black!50] (0,0) to (0,-2);
\draw[-,dashed,color=black!50] (0,0) to (1.5,-2); 
\draw[-,dashed,color=black!50] (0,0) to (2,-1.5); 

\draw[-] (-0.75,0) arc (180:217.5:0.75);
\node at (-0.5,0.2) {$\theta_1$};

\draw[-] (1.5,0) arc (0:-5.25:1.5);
\node at (1.3,0.2) {$\theta_2$};

\draw[-,dashed,color=black!25!lime,very thick] (1,-0.1) arc (-5.25:-90-52.5-1:1.005);

\node at (0,0) {\textcolor{red}{$\bullet$}};
\node at (0,0) {\textcolor{black}{$\circ$}};

\node at (1.5,-0.3) {\footnotesize\textcolor{black!25!lime}{$\mathcal{R}$}};
\node at (-1.075,-1) {\footnotesize\textcolor{teal}{$\mathcal{I}$}};

\draw[->,very thick,color=black!25!lime] (1,-0.1) to (2,-0.2);
\node at (1,-0.1) {\textcolor{black!25!lime}{$\bullet$}};

\draw[->,very thick,color=teal] (-0.81,-0.62) to (-2,-1.5);
\node at (-0.81,-0.62) {\textcolor{teal}{$\bullet$}};

\draw[-,dashed,color=black,very thick] (1.693,-0.1556) arc (-5.25:-90-52.5-1:1.7);
\end{tikzpicture}
\caption{Embedding of a KR braneworld with two positive tension branes in the black string geometry. The dashed black curve is the black string horizon separating the exterior and interior regions. The dashed green curve connecting the two branes is a putative RT surface.}
\label{pic:blackstring}
\end{centering}
\end{figure}
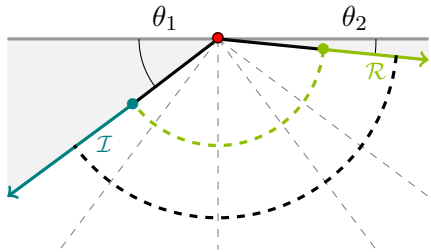

In the second case, the bulk geometry is the AdS$_5$ black string, which  replaces the metric on each constant-$\mu$ slice in Equ.~(\ref{metric1}) from empty AdS$_4$ by an AdS$_{4}$ planar black hole
\begin{equation}
    ds^2 = \frac{1}{u^2 \sin^2\mu}\left[-h(u) dt^2 + \frac{du^2}{h(u)} + d\vec{x}^2 + u^2 d\mu^2\right]\,,\label{eq:bstring}
\end{equation}
where the blackening factor is $h(u) = 1 - \frac{u^{3}}{u_h^{3}}$. The maximally extended geometry has two asymptotic boundaries (so we actually have two copies of the configuration in Fig.~\ref{pic:blackstring}). We computed the entanglement entropy between the two defect systems living on the separate asymptotic boundaries and we found that the entangling surface is just the black string horizon. This is consistent with the standard knowledge of an eternal black hole \cite{Maldacena:2001kr,Maldacena:2013xja} that the two holographic duals are in a thermofield double state with the entanglement entropy between them given by the bulk black hole horizon area. 

We emphasize that in both geometries Equ.~(\ref{metric1}) and Equ.~(\ref{eq:bstring}) constant-$\mu$ slices satisfies the junction equation Equ.~(\ref{eq:junction1}) with tensions given by
\begin{equation}
    T=(d-1)\cos(\mu)\,,
\end{equation}
where $d=4$ for bulk AdS$_5$ and the two Karch-Randall branes are just the $\mu=\theta_1$ and $\mu=\pi-\theta_2$ slices.

\subsection{The L/R Entanglement Entropy}
Moreover, in \cite{Geng:2020fxl} we defined and studied the so-called Left/Right (L/R) entanglement entropy that is naturally situated in wedge holography. It is defined as the entanglement entropy associated with an internal bipartition of the defect system into a left subsystem and a right subsystem. This definition is legitimate as the defect system is a nongravitating system and the associated Hilbert space can be factorized \cite{Geng:2020fxl}. The entanglement entropy associated with this bipartition can be calculated using the bulk Ryu-Takayanagi formula \cite{Geng:2020fxl}. 


This L/R entanglement entropy provides an important probe of the dynamics of the defect system. For example when the bulk geometry is a black hole, it is found that the L/R entropy obeys a unitary time-dependent Page curve \cite{Geng:2020fxl,Geng:2021iyq}. Interestingly, in a low dimensional case this L/R entanglement entropy can be studied even on the CFT side (the boundary description) in a slightly deformed setup and it is found to exactly match the result from the bulk calculation \cite{Geng:2021iyq}.

\section{Entanglement Entropy in 3D Wedge Holography}\label{sec:EE}
In this section, we generalize the above work to  bulk dimension  three (with classical Einstein's gravity in the bulk). In this case, when the bulk geometry is empty AdS$_3$ (see Fig.~\ref{pic:emptyads3}), comparing to $AdS_{5}$ (see Fig.~\ref{pic:emptyads5}), the spatial sector of the defect geometry degenerates from $S^2$ to two dots. Hence, the interpretation of the RT surface connecting the branes will now be the entanglement entropy between these two quantum dots. We will consider both cases that the two entangled dots are in a vacuum state and also in the thermofield double state. In the former case, the bulk geometry is as shown in Fig.~\ref{pic:blackstring} where we have two Karch-Randall branes living on an empty AdS$_3$ background. In the later case, the bulk geometry is a maximally extended BTZ black hole with two Karch-Randall branes forming a wedge (see Fig.~\ref{pic:ads3BTZ} for one side of the two-sided geometry).

\subsection{Vacuum State}\label{sec:vstate}
\begin{figure}
\centering
\begin{tikzpicture}

\draw[-,fill=gray!15,draw=none] (0-5,0-1) to (2-5,0-1) to (2-5,2-1) to (0-5,0-1);
\draw[-,fill=gray!15,draw=none] (0-5,0-1) to (-1.5-5,0-1) to (-1.5-5,2-1) to (0-5,0-1);

\draw[-,dashed] (-1.5-5,0-1) to (2-5,0-1);

\draw[-] (0.5-5,0-1) arc (0:45:0.5);

\node at (0.7-5,0.3-1) {$\theta_2$};

\draw[-] (-0.5-5,0-1) arc (180:127:0.5);

\node at (-0.7-5,0.3-1) {$\theta_1$};

\draw[-,thick,black] (0-5,0-1) to (-1.5-5,2-1);
\draw[-,thick,black] (0-5,0-1) to (2-5,2-1);

\node[red!100!] at (0-5,0-1) {$\bullet$};
\node at (0-5,0-1) {$\circ$};

\draw[-] (-7,-1.5) to (-7,1.5) to (-2.5,1.5) to (-2.5,-1.5) -- cycle;


\draw[-,black!40] (0,2) arc (90:-90:2);
\draw[-,black!40] (0,2) arc (90:270:2);

\draw[-,draw=none,fill=gray!15] (0,2) .. controls (-1.5,1) and (-1.5,-1) .. (0,-2) arc (-90:-270:2);
\draw[-,thick,black] (0,2) .. controls (-1.5,1) and (-1.5,-1) .. (0,-2);

\draw[-,draw=none,fill=gray!15] (0,2) .. controls (2,1) and (2,-1) .. (0,-2) arc (270:450:2);
\draw[-,thick,black] (0,2) .. controls (2,1) and (2,-1) .. (0,-2);

\node at (0,2) {\textcolor{blue!100!}{$\bullet$}};
\node at (0,2) {\textcolor{black}{$\circ$}};
\node at (0,-2) {\textcolor{red!100!}{$\bullet$}};
\node at (0,-2) {\textcolor{black}{$\circ$}};

\draw[-] (-0.25,-2.25) to (0.25,-2.25) to (0.25,-1.75) to (-0.25,-1.75) -- cycle;

\draw[-] (-0.25,-2.25) .. controls (-0.5,-2.5) and (-1.5,-2.5) .. (-2.5,-1.5);
\draw[-] (0.25,-2.25) .. controls (0,-3) and (-3.5,-3) .. (-7,-1.5);

\end{tikzpicture}
\caption{The empty AdS$_{3}$ wedge consisting of two KR branes. Here we show both the global picture and the Poincar\'e patch (which is one-point compactified to the global picture). While the defect in the Poincar\'e patch appears to be a point (which only has the time coordinate).}
\label{pic:emptyads3}
\end{figure}
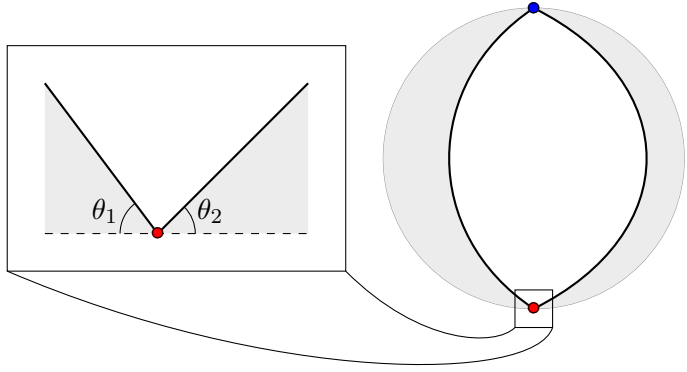

In parallel with Equ.~(\ref{metric1}), the AdS$_{3}$ metric in the Poincar\'{e} patch can be written as
\begin{equation}
    ds^2=\frac{1}{\sin^{2}\mu}[d\mu^2+\frac{-dt^2+du^2}{u^2}].\label{metric2}
\end{equation}
As we discussed in the previous section (see Fig.~\ref{pic:emptyads5}), to study the entanglement entropy of the defect system, we use the global patch for each AdS$_2$ slice with the metric then given by
\begin{equation}
    ds^2=\frac{1}{\sin^{2}\mu}\big[-\cosh^2(\eta)d\tau^2+d\eta^2+d\mu^2\big]\,,\label{metric3}
\end{equation}
where $\eta\in(-\infty,\infty)$, the two Karch-Randall branes are embedded as the $\mu=\theta_1$ and $\mu=\pi-\theta_2$ slices (see Fig.~\ref{pic:emptyads3}). Now the wedge holographic dual supported by the defect is a quantum mechanical system which only has the time dimension and it is \textit{conformally invariant}.

However, as we can see in Fig.~\ref{pic:emptyads3} that there are two such defects which live at $\eta\rightarrow\pm\infty$.  Unlike in the bulk five dimensional case, these two defects are now separated for topological reasons. We are interested in the entanglement entropy between these two defects. This can be calculated by looking for a one dimensional minimal surface in the 3d bulk that connects the two branes. Let the entangling surface be parameterized by $\eta(\mu)$ and $\tau=\text{const.}$, then the induced metric on this surface reads
\begin{equation}
    ds^{2}_{\text{induced}}=\frac{1}{\sin^2\mu}[1+\eta'(\mu)^2]d\mu^2\,,
\end{equation}
so the area (length) functional reads ($\theta_{1}$ and $\theta_{2}$ are defined as in Fig.~\ref{pic:emptyads3})
\begin{equation}
    A=\int_{\theta_1}^{\pi-\theta_2} d\mu \frac{1}{\sin\mu}\sqrt{1+\eta'(\mu)^2}.
\end{equation}
The variation of this area functional gives
\begin{equation}
    \delta A=\left.\frac{1}{\sin\mu}\frac{\eta'(\mu)}{\sqrt{1+\eta'(\mu)^2}}\delta\eta\right|_{\theta_1}^{\pi-\theta_2}-\int_{\theta_1}^{\pi-\theta_2} d\mu\delta\eta\frac{d}{d\mu}\big[\frac{1}{\sin\mu}\frac{\eta'(\mu)}{\sqrt{1+\eta'(\mu)^2}}\big]\,.\label{eq:variation}
\end{equation}
The solution for the resulting variational problem with both the bulk and boundary terms in Equ.~(\ref{eq:variation}) vanishing is $\eta'(\mu)=0$. Therefore the area of this entangling surface is
\begin{equation}
    A=\int_{\theta_1}^{\pi-\theta_2}\frac{d\mu}{\sin\mu}=\log(\cot{\frac{\theta_1}{2}})+\log(\cot{\frac{\theta_2}{2}})\,.\label{area}
\end{equation}
 Furthermore, using the Israel's junction equation \cite{Kraus:1999it} for the brane
 \begin{equation}
     K_{ab}=(K-T)h_{ab}\,,\label{junction}
 \end{equation}
we can relate the two angles $\theta_1, \theta_2$ to the tension of the two branes \cite{Fujita:2011fp,Karch:2020iit}
\begin{equation}
    T_{i}=\cos(\theta_{i})\,,\quad i=1,2\,.\label{tension}
\end{equation}
Thus, applying the Ryu-Takayanagi formula we get the entanglement entropy between the two defects as
\begin{equation}
    S_{\text{empty AdS$_3$}}=\frac{A}{4G_{3}}=\frac{1}{4G_3}\log(\sqrt{\frac{(1+T_1)(1+T_2)}{(1-T_1)(1-T_2)}})\,,\label{EE1}
\end{equation}
where $G_{3}$ is the bulk Newton's constant. 

Notice that the time coordinate $\tau$ of the geometry we are using (Equ.~(\ref{metric3})) is the same as the time coordinate in the global AdS$_3$ \cite{Karch:2020iit}. In global AdS$_3$ the metric is given by
\begin{equation}
    ds^2=-(r^2+1)d\tau^2+\frac{dr^2}{r^2+1}+r^2d\theta^2\,,
\end{equation}
where $r\in[0,\infty)$,$\theta\in(0,2\pi]$. Therefore the entanglement entropy Equ.~(\ref{EE1}) is also the entanglement entropy of a ground state of the Hamiltonian $\mathcal{H}_{\text{tot}}=\mathcal{H}_{1}+\mathcal{H}_{2}$\footnote{Notice that there is no interaction between these two quantum dots. This can be seen by thinking of dimensional reduction i.e. the theory that describes the two dots comes from a dimensional reduction of the CFT$_2$ which dualizes to the AdS$_3$ Einstein gravity (no Karch-Randall brane). This CFT$_2$ is a local theory.} that generates the time evolution in the global AdS$_3$, with $\mathcal{H}_1$ and $\mathcal{H}_2$ the Hamiltonians of the two defects respectively. This conclusion will be important for us to get the energy spectrum in the next section.

\subsection{Thermofield Double State}\label{sec:TFDstate}

In this subsection, we consider the thermofield double state of the Hamiltonian $\mathcal{H}_{\text{tot}}=\mathcal{H}_{1}+\mathcal{H}_{2}$ of the system of two decoupled defects as we noticed at the end of the previous subsection. 

By symmetry between the two defects, the two Hamiltonians $\mathcal{H}_{1}$ and $\mathcal{H}_{2}$ have the same spectrum $\{E_{0},E_{1},E_{2},\cdots\}$. The thermofield double state is given as
\begin{equation}
    \ket{\text{TFD($\beta$)}}=\frac{1}{\sqrt{Z}}e^{-\frac{\beta}{4}\mathcal{H}_{\text{tot}}}\sum_{n}\ket{E_{n}}\ket{E_{n}}=\frac{1}{\sqrt{Z}}\sum_{n}e^{-\frac{\beta E_{n}}{2}}\ket{E_{n}}\ket{E_{n}}\,.\label{eq:TFD}
\end{equation}
The dual bulk geometry is the BTZ black hole \cite{Maldacena:2001kr},
\begin{equation}
        ds^2=-(r^2-r_{h}^2)d\tau^2+\frac{dr^2}{r^2-r_{h}^2}+r^2 d\theta^2\,,\theta\in[0,2\pi)\,,\label{eq:sblackh}
\end{equation}
where there are two copies of the geometry in the maximally extended case with the asymptotic boundary at $r=\infty$ on each side (see Fig.~\ref{pic:ads3BTZ} for the one-sided geometry), $r_{h}$ is the radius of the black hole with the Hawking temperature given by $T_{H}=\frac{r_{h}}{2\pi}$ and we have two Karch-Randall branes intersecting at a defect on each of the asymptotic boundaries (see Fig.~\ref{pic:ads3BTZ}). 

However, in this case the configuration of the two Karch-Randall branes has to be found by explicitly solving the junction equation Equ.~(\ref{junction}) \cite{Fujita:2011fp}. Let's look at the left brane in Fig.~\ref{pic:ads3BTZ} which we call brane-$1$. We parametrize it as $\tau=\text{const.}$ and $\theta=\theta(r)$ which gives us the tangent vector of the brane in the $(\tau,r,\theta)$ coordinates
\begin{equation}
    t^{\mu}=(0,1,\theta'(r))\,,
\end{equation}
from which we can get the normalized normal vector
\begin{equation}
    n_{\mu}=\frac{(0,-\theta'(r),1)}{\sqrt{(r^2-r_{h}^{2})\theta'(r)^2+r^{-2}}}\,.
\end{equation}
Then solving the junction Equ.~(\ref{junction}) with brane tension $T=T_1$ we get
\begin{equation}
    \theta'(r)=\frac{T_{1}}{r\sqrt{r^2-T_{1}^2(r^2-r_{h}^2)}}
    \,,
\end{equation}
which integrates to 
\begin{equation}
    \theta(r)=\theta_{0}-\frac{1}{r_{h}}\arcsinh(\frac{T_{1}r_{h}}{r\sqrt{1-T_{1}^2}})\,,
\end{equation}
where $\theta_{0}$ is the position of the defect on the conformal boundary. Similarly, the right brane which we call brane-2 can be found as
\begin{equation}
    \theta(r)=\theta_{0}+\frac{1}{r_{h}}\arcsinh(\frac{T_{2}r_{h}}{r\sqrt{1-T_{2}^{2}}})\,.
\end{equation}
We assume that $T_{1}$, $T_{2}$ and $\theta_{0}$ are chosen such that neither of the branes intersects the perpendicular dashed black lines in Fig.~\ref{pic:ads3BTZ}.

The RT surface that we are looking for connects the two branes and it is time independent. For that purpose, we can focus on the geometry at $\tau=0$,
\begin{equation}
    ds^2|_{\tau=0}=\frac{dr^2}{r^2-r_{h}^2}+r^2d\theta^2\,.\label{eq:smetric0}
\end{equation}
For convenience, we can do the following coordinate coordinate transform
\begin{equation}
    z=\frac{1}{r}e^{-\theta r_{h}}\,,\quad x=\frac{1}{r_{h}}e^{-\theta r_{h}} \sqrt{1-\frac{r_{h}^2}{r^2}}\,,
\end{equation}
which transforms the metric Equ.~(\ref{eq:smetric0}) to
\begin{equation}
    ds^2|_{\tau=0}=\frac{dz^2}{z^2}+\frac{dx^2}{z^2}\,,\label{eq:smetric1}
\end{equation}
and the two branes to
\begin{equation}
    \begin{split}
         x_{L}^2+\left(z-\frac{e^{-\theta_{0}r_{h}}T_{1}}{r_{h}\sqrt{1-T_{1}^2}}\right)^2&=\frac{e^{-2\theta_{0}r_{h}}}{r_{h}^2(1-T_{1}^2)} \,,\\
    x_{R}^2+\left(z+\frac{e^{-\theta_{0}r_{h}}T_{2}}{r_{h}\sqrt{1-T_{2}^2}}\right)^2&=\frac{e^{-2\theta_{0}r_{h}}}{r_{h}^2(1-T_{2}^2)} \,.
    \end{split}
\end{equation}
Hence the spatial geometry is mapped to that in the Poincar\'{e} patch, the two branes are mapped to part of circles with the center in the bulk and out of the bulk respectively and both branes sit at $x^2=\frac{e^{-2\theta_{0}r_{h}}}{r_{h}^2}$ at the conformal boundary $z=0$. Moreover, the two asymptotic boundaries are now mapped to the single asymptotic boundary $z=0$ with the two defects at $x=\pm\frac{e^{-\theta_{0}r_{h}}}{r_{h}}$. For the sake of convenience, we define
\begin{equation}
    \begin{split}
        R_{1}&=\frac{e^{-2\theta_{0}r_{h}}}{r_{h}\sqrt{1-T_{1}^2}}\,,\quad \cos\theta_1=T_{1}\,,\\
        R_{2}&=\frac{e^{-2\theta_{0}r_{h}}}{r_{h}\sqrt{1-T_{2}^2}}\,,\quad \cos\theta_2=T_{2}\,,\label{eq:angletension}
    \end{split}
\end{equation}
and we know that 
\begin{equation}
    R_{1}\sin\theta_{1}=R_{2}\sin\theta_{2}=\frac{e^{-\theta_{0}r_{h}}}{r_{h}}\,.\label{eq:constraint0}
\end{equation}
We draw the brane configuration in the $(x,z)$ coodinates in Fig.~\ref{pic:easybrane} where $O$ is the origin $(0,0)$, $P_1$ and $P_2$ are the two defects $(\mp\frac{e^{\theta_{0}r_h}}{r_h},0)$ and $O_1$ and $O_2$ are the circular centers of the two branes respectively.

Now the RT surface is easy to find as discussed in \cite{Geng:2021iyq}. We parameterized the RT surface as $x=x(z)$ and its area functional reads
\begin{equation}
    A=\int_{z_{1}}^{z_{2}}\frac{dz}{z}\sqrt{x'(z)^2+1}\,,
\end{equation}
where $z_{1}$ and $z_{2}$ are the $z$ coordinates of the intersection of the RT surface with the two branes. This area functional gives an Euler-Lagrangian equation
\begin{equation}
    \frac{x'(z)}{z\sqrt{x'(z)^2+1}}=C\,,
\end{equation}
where $C$ is a constant and it can be determined by the position $z=z_{*}$ on the RT surface where the derivative $x'$ diverges as $C=\frac{1}{z_{*}}$. Assuming that the RT surface starts at $x(0)=x_0$ on the asymptotic boundary, this equation can be integrated to
\begin{equation}
    (x-x_{0}-z_{*})^2+z^2=z_{*}^2 \,,
\end{equation}
which is a circle whose center is at the conformal boundary $z=0$. 

As a summary, we have known that both the branes and the RT surface are circles in the geometry Equ.~(\ref{eq:smetric1}). Therefore, we can parameterize them as follows
\begin{equation}
    \begin{split}
        \text{$O_1$: }&x=R_{1}\sin\theta\,,\quad z=R_{1}(\cos\theta_{1}-\cos\theta)\,,\\
        \text{$O_2$: }&x=R_{2}\sin\tilde{\theta}\,,\quad z=R_{2}(\cos\tilde{\theta}-\cos\theta_{2})\,,\\
        \text{RT surface: }&x=x_{c}+R\sin\phi\,,\quad z=R\cos\phi\,,\label{eq:parametrization}
    \end{split}
\end{equation}
where $O_1$ and $O_2$ denote brane-1 and brane-2 (see Fig.~\ref{pic:easybrane}). 

The RT surface is fully characterized by the two parameters $x_{c}$ and $R$ and they can be determined from the boundary conditions of the RT surface when it ends on the branes. These boundary conditions have been worked out in \cite{Geng:2021iyq} by reparametrizing the RT surface as $x=x(s),z=z(s)$ using its normalized proper length $s\in[0,1]$. In this parametrization, the area functional reads
\begin{equation}
    A=\int_{0}^{1}ds\frac{1}{z}\sqrt{\dot{z}^2+\dot{x}^2} \,.
\end{equation}
With the fact that the Euler-Lagrange equation has been satisfied, so there is no bulk term of the area variation, the variation of the area functional becomes
\begin{equation}
    \delta A=\left.\frac{\dot{z}\delta{z}+\dot{x}\delta{x}}{z\sqrt{\dot{z}^2+\dot{x}^2}}\right|_{0}^{1} \,.
\end{equation}
Demanding that this $\delta A$ vanishes, we find that the end-points of the RT surface on the left and right branes obey
\begin{equation}
\begin{split}
    x'(z_{1})|_{\text{RT}}&=\frac{\dot{x}}{\dot{z}}|_{\text{RT}}=-\frac{\delta{z}}{\delta{x}}|_{\text{brane-1}z=z_{1}}=-\left.z'(x)\right|_{\text{brane-1},z=z_{1}} \,,\\
    x'(z_{2})|_{\text{RT}}&=\frac{\dot{x}}{\dot{z}}|_{\text{RT}}=-\frac{\delta{z}}{\delta{x}}|_{\text{brane-2},z=z_{2}}=-\left.z'(x)\right|_{\text{brane-2},z=z_{2}} \,,\label{eq:bc}
    \end{split}
\end{equation}
where $x'(z_{1})$ and $x'(z_{2})$ are the derivatives of the RT surface at its end-points. Using the parametrization of the branes and the RT surface in Equ.~(\ref{eq:parametrization}), we can see that at the point where the RT surface ends on the brane-1 we have
\begin{equation}
    \begin{split}
        R_{1}\sin\theta&=x_c+R\sin\phi_1\,,\\
        R_{1}(\cos\theta_{1}-\cos\theta)&=R\cos\phi_1\,\\
        -1&=(-\tan\phi_1)\tan\theta\,,\label{eq:bc1}
    \end{split}
\end{equation}
where $\phi_{1}$ is the value of the parameter $\phi$ for the ending point of the RT surface on the brane-1, the first two equations come from matching the $x$ and $z$ coordinates of the brane-1 and the RT surface and the third equation comes from the boundary condition Equ.~(\ref{eq:bc}). Combining these three equations we get
\begin{equation}
    x_{c}^2-R^2=R_{1}^2\sin^{2}\theta_{1}\,,\label{eq:constraint1}
\end{equation}
Similarly, we have the equations when the RT surface ends on the brane-2
\begin{equation}
    \begin{split}
        R_{2}\sin\tilde{\theta}&=x_c+R\sin\phi_2\,,\\
        R_{2}(\cos\tilde{\theta}-\cos\theta_{2})&=R\cos\phi_2\,\\
        -1&=\tan\phi_{2}\tan\tilde{\theta}\,,\label{eq:bc2}
    \end{split}
\end{equation}
which combine to
\begin{equation}
    x_{c}^2-R^2=R_{2}^2\sin^{2}\theta_{2}\,.\label{eq:constraint2}
\end{equation}

Therefore the two parameters of the RT surface $x_c$ and $R$ should be determined by solving the two equations Equ.~(\ref{eq:constraint1}) and Equ.~(\ref{eq:constraint2}). However, as we have noticed in Equ.~(\ref{eq:constraint0}) that we have $R_{1}\sin\theta_{1}=R_{2}\sin\theta_{2}$ so the two equations Equ.~(\ref{eq:constraint1}) and Equ.~(\ref{eq:constraint2}) actually degenerate and we only have one equation which constrains the two parameters $x_c$ and $R$. As a result, we have a one-parameter family of RT surfaces.

The area of a generic one of this one parameter family of RT surfaces can be computed as
\begin{equation}
    A=\int_{\phi_1}^{\phi_2}d\phi\frac{1}{\cos\phi}=\log\Big(\abs{\frac{1+\tan\frac{\phi_2}{2}}{1-\tan\frac{\phi_2}{2}}}\abs{\frac{1-\tan\frac{\phi_1}{2}}{1+\tan\frac{\phi_1}{2}}}\Big)\,.
\end{equation}
Solving Equ.~(\ref{eq:bc1}) and Equ.~(\ref{eq:bc2}) we get
\begin{equation}
    \tan\frac{\phi_1}{2}=-\frac{x_{c}+R_{1}}{R+R_{1}\cos\theta_{1}}\,,\quad\tan\frac{\phi_2}{2}=-\frac{x_{c}-R_{2}}{R+R_{2}\cos\theta_{2}}\,.
\end{equation}
Together with using Equ.~(\ref{eq:constraint1}) and Equ.~(\ref{eq:constraint2}) the area eventually simplifies to
\begin{equation}
    A=\log\big(\cot\frac{\theta_2}{2}\big)+\log\big(\cot\frac{\theta_1}{2}\big).
\end{equation}
Using Equ.~(\ref{eq:angletension}) and Equ.~(\ref{tension}), we finally get the area
\begin{equation}
    A=\log(\sqrt{\frac{1+T_2}{1-T_2}\frac{1+T_1}{1-T_1}})\,,
\end{equation}
which doesn't depend on the specific RT surface in the one-parameter family.

Therefore, we have a one-parameter family of RT surfaces which have equal area. Using the RT formula, according to which we take the minimum of the entangling surfaces, we get the entanglement entropy between the two defects ($P_1$ and $P_2$ in Fig.~\ref{pic:easybrane}) in the TFD state Equ.~(\ref{eq:TFD}) as 
\begin{equation}
    S_{\text{TFD}}=\frac{A}{4G_{3}}=\frac{1}{4G_3}\log(\sqrt{\frac{(1+T_1)(1+T_2)}{(1-T_1)(1-T_2)}})\,,\label{EE2}
\end{equation}
with $G_3$ the bulk Newton's constant. We notice that this entanglement entropy is the same as that in Equ.~(\ref{EE1}).
\begin{figure}
\begin{centering}
\begin{tikzpicture}[scale=1.5]
\draw[-,very thick,blue!40] (-3,0) to (3,0);
\node at (0,0) {\textcolor{red!100}{$\bullet$}};
\node at (0,0) {\textcolor{black}{$\circ$}};
\draw[-,dashed,very thick,black!40] (-3,-2) to (3,-2);
\draw[-,very thick,black] (0,0) arc (100:175:2.2); 
\draw[-,very thick,black] (0,0) arc (80:20:3.1); 
\draw[-,very thick,dashed,black] (-3,0) to (-3,-2);
\draw[-,very thick, dashed,black] (3,0) to (3,-2);
\draw[fill=gray, draw=none, fill opacity = 0.1] (-3,0)--(0,0) arc (100:175:2.2)--(-3,-2) ;
\draw[fill=gray, draw=none, fill opacity = 0.1] (3,0)--(0,0) arc (80:20:3.1)--(3,-2) ;
\draw[->,thick,red] (-2,0.5) to (2,0.5);
\node at (0,0.7){\textcolor{red}{$\theta$}};

\draw[-,very thick,blue!40] (-3,-3) to (3,-3);
\node at (0,-3) {\textcolor{blue!100}{$\bullet$}};
\node at (0,-3) {\textcolor{black}{$\circ$}};
\draw[-,dashed,very thick,black!40] (-3,-5) to (3,-5);
\draw[-,very thick,black] (0,-3) arc (100:175:2.2); 
\draw[-,very thick,black] (0,-3) arc (80:20:3.1); 
\draw[-,very thick,dashed,black] (-3,-3) to (-3,-5);
\draw[-,very thick, dashed,black] (3,-3) to (3,-5);
\draw[fill=gray, draw=none, fill opacity = 0.1] (-3,-3)--(0,-3) arc (100:175:2.2)--(-3,-5) ;
\draw[fill=gray, draw=none, fill opacity = 0.1] (3,-3)--(0,-3) arc (80:20:3.1)--(3,-5) ;
\draw[->,thick,red] (-2,-2.5) to (2,-2.5);
\node at (0,-2.3){\textcolor{red}{$\theta$}};
\end{tikzpicture}
\caption{\small{\textit{A typical brane configuration on a constant time $t=t_{0}$ slice on different sides of a bulk maximally extended BTZ black hole Equ.~(\ref{eq:sblackh}). The red and dark blue dots are the defects on each side and they are in a thermofield double state. The blue lines are the conformal boundary which goes from $0$ to $2\pi$ in $\theta$, the black curves are the two Karch-Randall branes, the dashed black horizontal lines are the black hole horizon $r=r_h$ and the shaded regions are removed (we assume brane tension positive). The whole geometry should be identified along the two vertical black dashed lines on each diagram.
We neglect the parts of the branes behind the black hole horizon which are easily seen to be timelike.}}}
\label{pic:ads3BTZ}
\end{centering}
\end{figure}
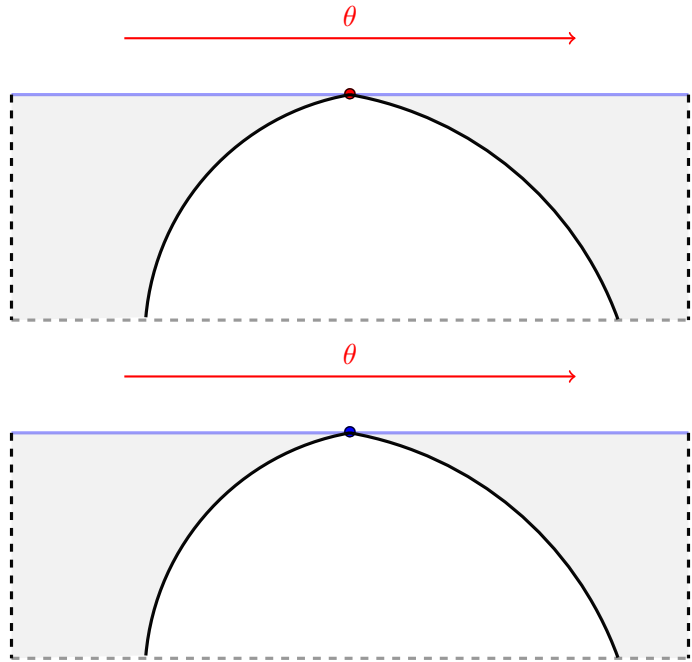
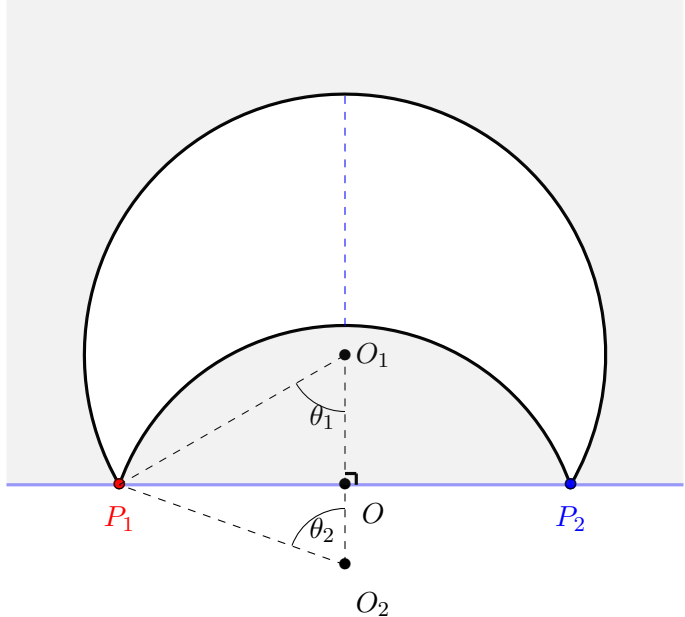
\begin{figure}
   \begin{centering}
   \begin{tikzpicture}[scale=1.5]
   \draw[-,very thick,blue!40] (-3,0) to (3,0);
   \draw[-,very thick,black] (-2,0) arc (210:-30:2.31);
   \draw[-,very thick,black] (-2,0) arc (160.5:19.5:2.12);
   \node at (-2,-0) {\textcolor{red!100}{$\bullet$}};
   \node at (-2,0) {\textcolor{black}{$\circ$}};
   \node at (2,0) {\textcolor{blue!100}{$\bullet$}};
   \node at (2,0) {\textcolor{black}{$\circ$}};
   \node at (0,1.15) {\textcolor{black}{$\bullet$}};
   \node at (0,-0.71) {\textcolor{black}{$\bullet$}};
   \node at (0.25,1.15) {\textcolor{black}{$O_{1}$}};
   \node at (0.25,-1.06) {\textcolor{black}{$O_{2}$}};
   \draw[-,dashed,black] (0,-0.71) to (0,1.15);
   \draw[-,dashed,black] (0,-0.71) to (-2,0);
   \draw[-,dashed,black] (0,1.15) to (-2,0);
   \draw[-] (0,-0.71+0.5) arc (90:160:0.5);
   \draw[-] (0,1.15-0.5) arc (-90:-150:0.5);
   \node at (-0.2,-0.4) {\textcolor{black}{$\theta_{2}$}};
   \node at (-0.2,0.6) {\textcolor{black}{$\theta_{1}$}};
   \draw[-,very thick,black] (0,0.1) to (0.1,0.1);
   \draw[-,very thick,black] (0.1,0.1) to (0.1,0);
   \node at (0,0) {\textcolor{black}{$\bullet$}};
   \node at (0.25,-0.25) {\textcolor{black}{$O$}};
   \node at (-2,-0.3) {\textcolor{red}{$P_1$}};
   \node at (2,-0.3) {\textcolor{blue}{$P_2$}};
   \draw[-,dashed,blue] (0,2.12-0.71) to (0,2.31+1.15);
   \draw[fill=gray, draw=none, fill opacity = 0.1] (-3,0)--(-2,0) arc (210:-30:2.31)--(3,0)--(3,4.3)--(-3,4.3);
   \draw[fill=gray, draw=none, fill opacity = 0.1] (2,0)--(-2,0) arc (160.5:19.5:2.12);
   \end{tikzpicture}
      \caption{The brane configuration in the $(x,z)$ coordinate. The shaded regions is the removed region in the bulk. The two black curves are the two branes and they are part of circles centered at $O_1$ and $O_2$. The dashed blue line is the image of the BTZ black hole horizon and we emphasize that at $t=0$ there is no black hole interior in our maximally extended geometry.}
   \label{pic:easybrane}
   \end{centering}
\end{figure}

\subsection{A Puzzle Regarding Entanglement Wedge Reconstruction}\label{sec:divergence}
Entanglement wedge reconstruction \cite{Jafferis:2015del,Dong:2016eik} sits at the center in the understanding of how the bulk geometry emerges from the boundary CFT data in the AdS/CFT correspondence. The basics statement of the entanglement wedge reconstruction that the bulk physics in the entanglement wedge of a boundary CFT subregion $\mathcal{A}$ denoted as EW($\mathcal{A}$)\footnote{EW($\mathcal{A}$) is defined as the bulk region bordered by $\mathcal{A}$ and its RT surface $\gamma(\mathcal{A})$.} is fully captured by the physics in $\mathcal{A}$. In other words, any operator in EW($\mathcal{A}$) can be written as an operator, or reconstructed by operators, in $\mathcal{A}$. 

An underlying assumption in the entanglement wedge reconstruction is that any boundary subregion $\mathcal{A}$ has a definite entanglement wedge and this is usually believed to be the case due to the minimization prescription in the RT formula. However, in many fine-tuned cases even in the standard AdS/CFT this is not the case. 

In this subsection, we will provide one such case in the 3d wedge holography and we will show later that this is due to the fact that the system is a trivial system and hence there shouldn't be any local bulk operator to reconstruct from the boundary.

In both cases for the entanglement entropy between the disconnected defects we considered in Sec.~\ref{sec:vstate} and Sec.~\ref{sec:TFDstate}, the above calculations show that there are an infinite number of degenerate RT surfaces.\footnote{Here we emphasize that this phenomenon of infinite number of degenerate RT surfaces was also recently noticed for different questions in other contexts---in de Sitter holography \cite{Geng:2019bnn} and in higher dimensional Karch-Randall braneworld \cite{Geng:2020fxl}.  In both cases it is understood that this phenomenon is due to a continuous symmetry of the system. As we will see, this is also the case in the current situation.} In the case that the bulk is empty AdS$_3$, the entangling surfaces are given by
\begin{equation}
    \eta'(\mu)=0\,.
\end{equation}
Thus the RT surfaces are just $\eta=C$ for any real constant $C$. In the case that the bulk is a BTZ black hole, there are two parameters $x_c$ and $R$ that characterize the RT surfaces but they only have to satisfy one constraint Equ.~(\ref{eq:constraint1}) so we have a one-parameter family of RT surfaces. Our calculations showed that in both cases the area of the RT surfaces are the same.

In the original Ryu-Takayanagi proposal for the entanglement entropy \cite{Ryu:2006bv,Ryu:2006ef}, when we have more than one candidate entangling surface the path integral is dominated by the one with the smallest area, which is the relevant entangling surface. Applying this prescription we would have a well-defined entanglement entropy between the two defects in both cases. However, the infinite degeneracy of the RT surfaces says that there is no definite entanglement wedge for either one of the defects. This naively imposes a challenge for the \textit{entanglement wedge reconstruction}. Nevertheless, one possible interpretation of this puzzle is that the holographic dual of the wedge is a trivial system with trivial dynamics and hence only ground states. In the language of the bulk, there is no perturbative local bulk operator in the 3d wedge. In other words, there is no probe limit for bulk objects and their backreaction on the geometry will destroy the wedge setup which is indeed the case by \cite{Geng:2021iyq}.\footnote{There we showed that the branes will generically intersect in the bulk.}

We will show more precisely in Sec.~\ref{sec:spectrumEE} that the above speculation is indeed the case for us. We will compute the exact energy spectrum of the dual defect system and see that it only contains degenerate ground states with no excitations. The conformal symmetry plays an important role in this calculation. Moreover, as we will see in Sec.~\ref{sec:confbreaking}, that taking into account the brane fluctuations will lift this RT surfaces degeneracy.
\subsection{Further Evidence for the Triviality}
Before we turn to Sec.~\ref{sec:spectrumEE} for an explicit demonstration that the system is trivial with no excitations, we will provide further evidence for this speculation in this subsection.

\subsubsection{The L/R Entanglement Entropy}\label{sec:L/Rtrivialdyn}
Let's firstly study the L/R entanglement entropy in this low dimensional wedge holographic setup which captures certain aspects of the dynamics of the system. As 
it is discovered in \cite{Geng:2020qvw} that there are four candidate entangling surfaces in the higher dimensional case---the finite island surface ending on the left brane, the finite island surface ending on the right brane, the small island surfaces and the Hartman-Maldacena surface. The first two types of surfaces connect the defect to a point away from the defect on the left and right branes respectively. The third type of surfaces connect the defect to a point infinitesimally close to the defect and this point lives either on the left or right brane. The fourth type of surface goes through the bulk and connects the defects on the opposite asymptotic boundaries. However, in the 3d case we only have two candidates---the Hartman-Maldacena surface and the small island surface. The reason is as follows.

In the case of the vacuum AdS$_3$ as we discussed in Sec.~\ref{sec:vstate} the brane ending RT surfaces satisfy $\eta'(\mu)=0$ or equivalently $\eta=\text{const}.$ so for the RT surface shot from the defect which is $\eta=\infty$ it would be $\eta=\infty$. Therefore, in the language of \cite{Geng:2020fxl}, the brane ending RT surfaces is the small island RT surface and the finite RT surface doesn't exist in the current context. The area of this small island RT surface now reads
\begin{equation}
    A=\int_{\theta_{1}}^{\pi}\frac{d\mu}{\sin\mu}=-\log(\cot\frac{\theta_{\epsilon}}{2})+\log(\cot\frac{\theta_{1}}{2})=\log(\frac{2a}{\epsilon})+\log\cot\frac{\theta_{1}}{2}\,,\label{eq:islandsur}
\end{equation}
where $\theta_{\epsilon}$ is very close to $\pi$ and in the last step we transformed back to the Poincare patch that $z=u\sin\mu$ where near the boundary we have $\epsilon=a\sin\theta_{\epsilon}$ for $a$ as the u-coordinate of the anchor point. As take the anchor point to the defect we would have $a\rightarrow0$.


For the thermofield double state or the BTZ black hole we have  similar reasoning. As we have seen in Sec.~\ref{sec:TFDstate} (Fig.~\ref{pic:easybrane}) the branes are part of circles with centers in or out of the bulk and the RT surfaces ending on the branes are also circular with its diameter living along the asymptotic boundary. As a result, for the RT surfaces shooting from the defects that end on the brane with the correct boundary condition Equ.~(\ref{eq:bc}) (i.e. the RT surface ends on the brane perpendicularly) it must be an infinitesimal semicircle just as in the case of vacuum AdS$_3$. Therefore, we also have the fact that in the thermofield double state or the BTZ black hole the finite island surfaces degenerate with the small island surface for the L/R entropy. With the same reasoning as in the vacuum AdS$_3$ case, the entanglement entropy calculated by such entangling surfaces is minus infinity.

Though in both cases we still have the Hartman-Maldacena surface which connects the two disconnected defects in both the empty AdS$_3$ case and the BTZ black hole case. As it is standard in the holographic calculation of entanglement entropy, the entanglement entropy calculated by this surface is (positive) infinitely divergent (with no $\ln a$ suppression as for island surface in Equ.~(\ref{eq:islandsur})). Hence, using the RT prescription we see that the L/R entanglement entropy in AdS$_3$ wedge holography is always captured by the small island surface and its value is a constant for both the vacuum state  and for thermofield double state.

This is in contrast with the result we found in the higher dimensional cases \cite{Geng:2020fxl}. Moreover, in the language of \cite{Geng:2020fxl} we can say that in the 3d case the bipartite Hilbert space dimension of the defect system is not large enough to see the time-dependence of the L/R entanglement entropy.

To be precise, here we give the formula for HM surface and small island surface contributions to the L/R entropy in the BTZ case. We refer the readers to \cite{Geng:2021hlu} for details of the calculation.
\begin{equation}
    S_{\text{L/R}}=\min\Biggl[\lim_{a\rightarrow0}\frac{c}{3}\ln(\frac{a}{\epsilon})+2\ln(g_{L}),\,\frac{c}{3}\ln(\frac{\beta}{\pi\epsilon}e^{-\frac{2\pi\theta_{0}}{\beta}}\cosh\left(\frac{2\pi t}{\beta}\right)),\,\lim_{a\rightarrow0}\frac{c}{3}\ln(\frac{a}{\epsilon})+2\ln(g_{R})\Biggr],
    \label{eq:entropy-bh-gravity}
\end{equation}
where $g_{L}$ and $g_{R}$ are respectively the boundary entropy associated with the left and right branes and $\theta_{0}$ is the position of the defect on the asymptotic boundary. As we can see that for finite $g_{L}$ and $g_{R}$ the time dependent candidate is always subdominant and $S_{\text{L/R}}$ is captured by the small island RT surfaces and stays as a constant in time. This is consistent with our suggestion that the system has trivial dynamics.

\subsubsection{The 2d Effective Action and the Entanglement Entropy}\label{sec:puzzle2}
 In this subsection, we will derive the effective action of the localized gravity on the Karch-Randall branes and we will see that this action can be used to reproduce our results on the entanglement entropy including the degeneracy of the RT surfaces. Moreover, this gravitational action is purely topological containing trivial dynamics. 


We start from the metric Equ.~(\ref{metric3}) but with a coordinate change $\mu\rightarrow r$
\begin{equation}
    r=\log(\cot\frac{\mu}{2})\,,
\end{equation}
which transforms the metric to
\begin{equation}
\begin{split}
    ds^2&=dr^2+\cosh^{2}(r)\big[-\cosh^{2}(\eta)d\tau^2+d\eta^2\big]\,.\label{metric4}
    \end{split}
\end{equation}
In this metric, a Karch-Randall brane is a constant $r$ slice with tension given by
\begin{equation}
    T=\tanh{\abs{r}}\,.
\end{equation}
We have two positive tension branes at $r=r_{1}<0$ and $r=r_{2}>0$ with tensions $T_{1}=\tanh(-r_{1})$ and $T_{2}=\tanh{r_{2}}$. 

To obtain the effective action for localized gravity we consider the following metric
\begin{equation}
    ds^2=dr^2+\cosh^{2}(r)g_{\mu\nu}(x)dx^{\mu}dx^{\nu}\,,
\end{equation}
where $g_{\mu\nu}(x)$ is the 2d dynamical metric on the brane. This is actually the would-be massless graviton mode built upon which there is a tower of massive graviton modes \cite{Karch:2000ct}.\footnote{Here the (would-be) massless graviton mode is normalizable (due to the presence of two Karch-Randall branes) so it is the only geometrical mode at low energy. This can be understood by the standard AdS/CFT duality which says that the massless mode sources the energy-momentum tensor and the massive modes source the descendants of the energy-momentum tensor. In our case, we expect that the (2d) energy-momentum tensor is trivial as the system has trivial dynamics. Hence, here we expect that those massive modes, which correspond to the descents of the trivial energy-momentum tensor, would also be trivial. This is confirmed later by the fact that the entanglement entropy between the two disconnected defects calculated by the effective action of the (would-be) massless graviton mode precisely reproduces the calculation from the bulk.} Using this metric ansatz the 3d Ricci scalar can be calculated as
\begin{equation}
    R[g_{\text{bulk}}]=\frac{1}{\cosh^{2}r}R[g]-4-2\tanh^{2}r\,.
\end{equation}
Then the effective action is obtained by evaluating the bulk Einstein-Hilbert action with the appropriate boundary terms on the branes
\begin{equation}
            \begin{split}
        S_{\text{3d}}=&-\frac{1}{16\pi G_3}\int d^{3}x \sqrt{-g_{\text{bulk}}}(R[g_{\text{bulk}}]+2)-\frac{1}{8\pi G_{3}}\int_{i=1,2} d^{2}x\sqrt{-h_{i}}(K_{i}-T_{i})\\
        =&-\frac{1}{16\pi G_3}\int d^{3}x \sqrt{-g_{\text{bulk}}}(R[g_{\text{bulk}}]+2)-\frac{1}{8\pi G_{3}}\int_{i=1,2} d^{2}x\sqrt{-h_{i}}\cosh^{2}(r_{i})(K_{i}-T_{i})\\
        =&-\frac{1}{16\pi G_3} \int_{r_{1}}^{r_{2}}dr\int d^{2}x \sqrt{-g}\Big[R[g]-2\big(\cosh^2(r)+\sinh^{2}(r)\big)\Big]\,\\
        &-\frac{1}{8\pi G_{3}}\int d^{2}x\sqrt{-g}\cosh^{2}(r_{1})\tanh(-r_1)-\frac{1}{8\pi G_{3}}\int d^{2}x\sqrt{-g}\cosh^{2}(r_{2})\tanh(r_{2})\,\\
        =&-\frac{r_{2}-r_{1}}{16\pi G_3}\int d^{2}x \sqrt{-g}R[g]\,,\label{eq:EFT}
    \end{split}
\end{equation}
where we used the fact that we have tensions $T_{1}=\tanh(-r_{1})$ and $T_{2}=\tanh(r_{2})$, the induced metric $h_{\mu\nu}=\cosh^{2}(r)g_{\mu\nu}$ and the Israel's junction condition \cite{Israel:1967zz}
\begin{equation}
    K_{\mu\nu}=T\space h_{\mu\nu}\,\implies\quad K=2T\,.
\end{equation}

As a result, we get the following effective action
\begin{equation}
    S_{eff}=-\frac{r_{2}-r_{1}}{16\pi G_{3}}\int d^{2}x\sqrt{-g}R[g]\,.\label{eq:2d}
\end{equation}
It deserves to be mentioned that this is precisely the model considered by Marold and Maxfield in \cite{Marolf:2020xie}\footnote{We thanks Douglas Stanford to bring this paper to our attention.} but here we consider the Lorenzian version where we have the usual duality between a bulk geometry and a boundary state and can study holographic entanglement entropy. On the one hand, this action is purely topological with no dynamics. On the other hand, this is reminiscent of Jackiw-Teitelboim gravity \cite{Teitelboim:1983ux,Jackiw:1984je,Maldacena:2016upp} but with a constant dilaton field.\footnote{That is the AdS$_2$ isometry is not broken.} From this action we have the entropy functional for the entanglement entropy between the two asymptotic boundaries of the AdS$_2$ (i.e. the two defects) 
\begin{equation}
    S=\frac{r_{2}-r_{1}}{4 G_3}\,.
\end{equation}
Using the brane tensions 
\begin{equation}
    T_{1}=\tanh(-r_{1})\,\quad,T_{2}=\tanh(r_{2})\,,
\end{equation}
we get
\begin{equation}
    S=\frac{1}{4G_{3}}\log(\sqrt{\frac{(1+T_1)(1+T_2)}{(1-T_1)(1-T_2)}})\,.
\end{equation}
However, similar to our 3d analysis in Sec.\ref{sec:divergence}, there is an infinite degeneracy of RT surfaces due to the fact that the area functional is a constant or the 2d dilaton field is a constant. That is there are in fact an infinity number of degenerate saddle points over a constant time slice of the 2d AdS$_2$ manifold for the following minimization problem
\begin{equation}
    S=\min_{y}(\frac{\phi(y)}{4G_{2}})\,,
\end{equation}
where $y$ is the coordinate on a constant time slice of AdS$_2$, $G_{2}$ is the induced 2d Newton's constant $\frac{r_{2}-r_{1}}{G_{3}}$ and in our case the dilaton field $\phi(y)=1$ is a constant.

In summary, we have reproduced our results for the entanglement entropy in Sec.~\ref{sec:EE}, both of its value and the infinite degeneracy of the RT surfaces, from the 2d effective action of the localized gravity Equ.~(\ref{eq:2d}). The fact that the effective action has no dynamics is consistent with our expectation that the system is trivial and has no perturbative excitations. Furthermore, we notice that the fact that the 2d dilaton field is a constant, or equivalently the conformal symmetry, plays a deterministic role for the infinite degeneracy of the RT surfaces.

\section{From Entanglement Entropy to the Energy Spectrum}\label{sec:spectrumEE}
In this section, we fully exploit the effect of the conformal symmetry. We will see that the calculations of the entanglement entropy between the two disconnected defects in the previous section together with the conformal invariance of the dual quantum mechanics system gives us the explicit form of the energy spectrum.

For a generic (unitary) conformal quantum mechanics system, there is no scale.  Therefore, by dimensional analysis, the energy density is fixed up to two unknown real (positive) constants $\alpha$ and $\gamma$ as
\begin{equation}
    \rho(E)=\alpha\delta(E)+\frac{\gamma}{E}.
\end{equation}
In the cases that we considered in the previous section, we have two such systems as the two disconnected defects and they are not interacting. Interestingly, we notice that the entanglement entropy between these two systems in a vacuum state Equ.~(\ref{EE1}) is the same as that when they are in a TFD state Equ.~(\ref{EE2}). Moreover, in the TFD state Equ.~(\ref{eq:TFD}) the entanglement entropy Equ.~(\ref{EE2}) is independent of the inverse temperature $\beta$. Hence, we can see that the energy spectrum only contains degenerate ground states or it is infinitely gapped i.e.
\begin{equation}
    \gamma=0,
\end{equation}
and the ground state degeneracy $\alpha$ is given by the entanglement entropy as
\begin{equation}
    \alpha=e^{S_{\text{TFD}}}=e^{S_{\text{empty AdS$_3$}}}=\Bigg[\frac{(1+T_1)(1+T_2)}{(1-T_1)(1-T_2)}\Bigg]^{\frac{1}{8G_3}}.
\end{equation}

In summary, we get the energy density of the wedge holographic dual of AdS$_3$ Einstein's gravity as
\begin{equation}
    \rho(E)=\Bigg[\frac{(1+T_1)(1+T_2)}{(1-T_1)(1-T_2)}\Bigg]^{\frac{1}{8G_3}}\delta(E).\label{eq:energy}
\end{equation}
We emphasize that this should be understood as the energy spectrum of the dual quantum mechanics system of a UV-complete AdS$_2$ quantum gravity theory induced on the Karch-Randall brane from the Einstein's gravity in AdS$_3$ bulk. As is well known there are no nonzero energy states for UV-complete quantum gravity in AdS$_2$ \cite{Maldacena:1998uz} that preserves the AdS$_2$ asymptotic symmetry (of which the AdS$_2$ isometry is a finite subgroup).\footnote{This can be seen straightforwardly by considering the energy-momentum tensor of the boundary description. The boundary theory should be (locally) conformally invariant so its energy-momentum tensor is traceless. However, as a 1d system its energy-momentum tensor only has one component---the energy. Hence the energy is constantly zero for any states or it only contains zero energy states.} Hence our result is consistent with this expectation.

\section{Setting the Brane Free---Emergence of Dilaton Gravity}\label{sec:effactionJT}
In the previous sections, we focused on the physics of Einstein's gravity in a wedge formed by two rigid Karch-Randall branes in an AdS$_3$ bulk. In the case with just one brane, the rigidity of the brane is usually claimed to be achieved by putting the brane at an orbifold fix point similar to an O-plane in string theory. However, if we remind ourself of the usual Randall-Sundrum scenario I \cite{Randall:1999ee,Randall:1999vf} there would be a scalar mode called a radion which describes the brane fluctuations. In the Randall-Sundrum language, this mode controls the UV/IR hierarchy and hence has important phenomenological implications. In our case, we will see in this section that, with the brane fluctuations carefully taken into account, it enables us to embed a type of dilaton gravity theories into the Karch-Randall braneworld. In particular, the Jackiw-Teitelboim gravity \cite{Jackiw:1984je,Teitelboim:1983ux} is a special example.

\subsection{General Situations}
In this subsection, we consider the general situations that we have two Karch-Randall branes with generic tensions $T_{1}$ and $T_{2}$. The AdS$_3$ gravitational action with appropriate boundary terms on these two Karch-Randall branes takes the following form
\begin{equation}
\begin{split}
    S=-\frac{1}{16\pi G_3}\int d^3 x\sqrt{-g_{\text{bulk}}}\Big[R[g_{\text{bulk}}]+2\Big]-\frac{1}{8\pi G_3}\int_{\text{brane1}\cup\text{brane2}}d^2 x \sqrt{-h}(K-T)\,,\label{eq:action}
    \end{split}
\end{equation}
where we have set the curvature length to be one and $h_{\mu\nu}$ is the induced metric on the brane.

The metric of empty AdS$_3$ can be written in the form
\begin{equation}
    ds^2=dr^2+\cosh^{2}(r)\Big(-\cosh^2(\eta)dt^2+d\eta^2\Big)\,,
\end{equation}
which is foliated by AdS$_2$ slices at constant $r$. The two branes brane1 and brane2 are just the $r=r_1$ and $r=r_2$ slices. Solving the Israel's junction equation we get
\begin{equation}
    T_{1}=\frac{1}{2}K_{1}=\tanh(-r_{1})\,,\quad T_{2}=\frac{1}{2}K_{2}=\tanh(r_2)\,,
\end{equation}
where we take $r_{1}<0$ and $r_{2}>0$. 

Now we consider the dynamics of this two-brane system such that the brane tensions are fixed but the branes' internal shape and their embedding in the ambient 3-d space are fluctuating. Therefore, we consider the following 3-d metric
\begin{equation}
    ds^2=dr^2+\cosh^{2}(r)g_{\mu\nu}(x)dx^{\mu}dx^{\nu}\,,\label{eq:ansatz}
\end{equation}
where we gauge fix the bulk diffeomorphism to eliminate $g_{\mu r}$ and set $g_{rr}=1$ and we consider the low energy dynamics where $g_{\mu\nu}$ has no $r$-dependence. We let the two branes to be embedded as $r_{1}(x)=r_{1}+\delta\phi_{1}(x)$ and $r_{2}(x)=r_{2}+\delta\phi_{2}(x)$ where $x$ denotes the two-dimensional coordinates on the brane.

To get the effective action describing the low-energy dynamics of this system, we have to plug our ansatz into the action Equ.~(\ref{eq:action}). As we saw before, the bulk Ricci scalar of the metric Equ.~(\ref{eq:ansatz}) can be calculated as
\begin{equation}
    R[g_{\text{bulk}}]=\frac{1}{\cosh^{2}r}R[g]-4-2\tanh^{2}(r)\,.
\end{equation}
If we care only about our effective action to the first order in $\delta\phi_{1}(x)$ and $\delta\phi_{2}(x)$, we don't have to consider derivative terms in $\delta\phi_{1}(x)$ and $\delta\phi_{2}(x)$ as they are just total derivatives. Therefore, we can use the following approximation
\begin{equation}
    K_{1}=2\tanh(-r_{1}-\delta\phi_{1}(x))\,,\quad K_{2}=2\tanh(r_{2}+\delta\phi_{2}(x))\,.
\end{equation}

As a result, we can calculate the effective action:
\begin{equation}
    \begin{split}
        S_{\text{eff,I}}=&-\frac{1}{16\pi G_{3}}\int_{r_{1}+\delta\phi_{1}}^{r_{2}+\delta\phi_{2}}dr\int d^{2}x\sqrt{-g}\Big[R[g]-4\cosh^{2}(r)-2\sinh^{2}(r)+2\cosh^{2}(r)\Big]\,\\
        &-\frac{1}{8\pi G_{3}}\int d^{2}x\sqrt{-g}\cosh^{2}(r_{1}+\delta\phi_{1})\Big(2\tanh(-r_{1}-\delta\phi_{1})-\tanh(-r_{1})\Big)\,\\
        &-\frac{1}{8\pi G_{3}}\int d^{2}x\sqrt{-g}\cosh^{2}(r_{2}+\delta\phi_{2})\Big(2\tanh(r_{2}+\delta\phi_{2})-\tanh(r_{2})\Big)\,\\
        =&-\frac{r_{2}-r_{1}}{16\pi G_{3}}\int d^{2}x\sqrt{-g}R[g]-\frac{1}{16\pi G_3}\int d^{2}x\sqrt{-g}(\delta\phi_{2}-\delta\phi_{1})\Big[R[g]+2\Big]\,.
    \end{split}
\end{equation}
where we notice that the first term is the same as in Equ.~(\ref{eq:EFT}) for which the branes are not fluctuating. If we define $\phi(x)=\delta\phi_{2}(x)-\delta\phi_{1}(x)$, then we get the Jackiw-Teitelboim gravity 
\begin{equation}
    S_{\text{eff,I}}=-\frac{r_{2}-r_{1}}{16\pi G_{3}}\int d^{2}x\sqrt{-g}R[g]-\frac{1}{16\pi G_{3}}\int d^{2}x\sqrt{-g}\phi(x)\Big[R[g]+2\Big]\,.
    \label{eq:JT1}
\end{equation}
However, if one goes to the quadratic order in $\delta\phi_{1}$ and $\delta\phi_{2}$ one would have the following additional action
\begin{equation}
\begin{split}
    S_{\text{dilaton}}=-\frac{1}{8\pi G_{3}}\int d^{2}x\sqrt{-g}\Big[&\frac{\tanh(r_{2})}{2}\nabla_{\mu}\delta\phi_{2}\nabla^{\mu}\delta\phi_{2}+\tanh(r_{2})(\delta\phi_{2})^2\\&-\frac{\tanh(r_{1})}{2}\nabla_{\mu}\delta\phi_{1}\nabla^{\mu}\delta\phi_{1}-\tanh(r_{1})(\delta\phi_{1})^2\Big]\,.\label{eq:dilaton2d}
    \end{split}
\end{equation}

The full low energy effective action would be up to $\mathcal{O}(\delta\phi^2)$ and would be the sum of Equ.~(\ref{eq:JT1}) and Equ.~(\ref{eq:dilaton2d}). Moreover, the full action can be simplified by introducing
\begin{equation}
    \psi(x)=\frac{T_{1}}{\sqrt{T_{1}+T_{2}}}\delta\phi_{1}(x)+\frac{T_{2}}{\sqrt{T_{1}+T_{2}}}\delta\phi_{2}(x)\,,
\end{equation}
by which the dilaton part simplifies as
\begin{equation}
    S_{\text{dilaton}}=-\frac{1}{8\pi G_{3}}\int d^{2}x\sqrt{-g}\Bigg[\frac{T_{1}T_{2}}{2(T_{1}+T_{2})}\nabla_{\mu}\phi(x)\nabla^{\mu}\phi(x)+\frac{T_{1}T_{2}}{T_{1}+T_{2}}\phi^{2}(x)+\frac{1}{2}\nabla_{\mu}\psi(x)\nabla^{\mu}\psi(x)+\psi^{2}(x)\Bigg]\,.
\end{equation}
A Weyl transform of the 2d metric
\begin{equation}
    g_{\mu\nu}(x)\rightarrow e^{-\frac{T_{1}T_{2}\phi(x)}{T_{1}+T_{2}}}g_{\mu\nu}(x)\,,
\end{equation}
can be used to eliminate the kinetic and mass terms of $\phi(x)$ to order $\mathcal{O}(\phi^{2})$.

As a result, we have the full effective action
\begin{equation}
\begin{split}
    S_{\text{eff}}=-\frac{r_{2}-r_{1}}{16\pi G_{3}}\int d^{2}x\sqrt{-g}R[g]-\frac{1}{16\pi G_{3}}\int d^{2}x\sqrt{-g}\phi(x)\Big[R[g]+2\Big]\\-\frac{1}{8\pi G_{3}}\int d^{2}x\sqrt{-g}\Bigg[\frac{1}{2}\nabla_{\mu}\psi(x)\nabla^{\mu}\psi(x)+\psi^{2}(x)\Bigg]\,,
    \label{eq:JT1final}
    \end{split}
\end{equation}
which is a type of 2d dilaton gravity. More precisely, it is  Jackiw-Teitelboim gravity minimally coupled to a massive scalar field.\footnote{We refer interested readers to \cite{Achucarro:1993fd,Mertens:2018fds,Harlow:2018tqv,Ghosh:2019rcj,Maxfield:2020ale,Verheijden:2021yrb,Numasawa:2022cni,Suzuki:2022xwv,Eberhardt:2022wlc,Kusuki:2022wns} for other speculations and investigations of the relationship between JT gravity and AdS$_3$ Einstein's gravity and see \cite{Nayak:2018qej,Yang:2018gdb,Sachdev:2019bjn,Maldacena:2019cbz,Heydeman:2020hhw,Iliesiu:2020qvm,Moitra:2022glw,Boruch:2022tno,Gukov:2022oed} for the emergence of JT gravity in other context.}

\subsection{Two Special Cases---The Emergence of Pure JT Gravity}
In this subsection, we provide two special cases where the 2d dilaton gravity we have obtained is  pure JT gravity. The first case is a straight limit of Equ.~(\ref{eq:dilaton2d}) and in the second case we project out one of the dilatons by orbifolding our theory.
\subsubsection{Tensionless Branes}
We can see from Equ.~(\ref{eq:dilaton2d}) that the dilaton kinetic terms $S_{\text{dilaton}}$ of the effective action vanishes if the two branes are initially tensionless. Hence we see that the pure JT gravity describes the low energy gravitational dynamics for the tensionless branes. Interestingly, in this case the wedge in the AdS$_3$ bulk degenerates if we don't have the fluctuation of the branes. Therefore the Jackiw-Teitelboim gravity captures the emergence of the extra dimension to the brane.

\subsubsection{Orbifolding}
\label{sec:orbifolding}
In the original work of Karch and Randall \cite{Karch:2000ct} there is only one brane. The brane can be thought of as living at an orbifold fixed point between two mirrorly symmetric universes in order for us to safely ignore the scalar mode describing the brane fluctuations in the ambient space. However, in our current situation, with two branes, we have to start with taking into account the fluctuations of both branes and comparing to the Randall-Sundrum I scenario \cite{Randall:1999ee} the orbifolding procedure is nontrivial in the generic case, where the two branes have different tension, as we don't have an obvious $Z_{2}$ symmetry to orbifold our theory.

\begin{figure}
\centering
\begin{tikzpicture}
\draw[-,black!40] (0,1.5) arc (90:-90:1.5);
\draw[-,black!40] (0,1.5) arc (90:270:1.5);
\draw[-,draw=none,fill=gray!15] (0,1.5) .. controls (-1.2,0.8) and (-1.2,-0.8) .. (0,-1.5) arc (-90:-270:1.5);
\draw[-,thick,green] (0,1.5) .. controls (-1.2,0.8) and (-1.2,-0.8) .. (0,-1.5);

\draw[-,draw=none,fill=gray!15] (0,1.5) .. controls (1.5,0.8) and (1.5,-0.8) .. (0,-1.5) arc (270:450:1.5);
\draw[-,thick,orange] (0,1.5) .. controls (1.5,0.8) and (1.5,-0.8) .. (0,-1.5);
\node at (0,1.5) {\textcolor{blue!100!}{$\bullet$}};
\node at (0,1.5) {\textcolor{black}{$\circ$}};
\node at (0,-1.5) {\textcolor{red!100!}{$\bullet$}};
\node at (0,-1.5) {\textcolor{black}{$\circ$}};

\draw[-,black!40] (3.2,1.5) arc (90:-90:1.5);
\draw[-,black!40] (3.2,1.5) arc (90:270:1.5);
\draw[-,draw=none,fill=gray!15] (3.2,1.5) .. controls (4.4,0.8) and (4.4,-0.8) .. (3.2,-1.5) arc (-90:90:1.5);
\draw[-,thick,green] (3.2,1.5) .. controls (4.4,0.8) and (4.4,-0.8) .. (3.2,-1.5);
\draw[-,draw=none,fill=gray!15] (3.2,1.5) .. controls (1.7,0.8) and (1.7,-0.8) .. (3.2,-1.5) arc (-90:-270:1.5);
\draw[-,thick,orange] (3.2,1.5) .. controls (1.7,0.8) and (1.7,-0.8) .. (3.2,-1.5);
\node at (3.2,1.5) {\textcolor{blue!100!}{$\bullet$}};
\node at (3.2,1.5) {\textcolor{black}{$\circ$}};
\node at (3.2,-1.5) {\textcolor{red!100!}{$\bullet$}};
\node at (3.2,-1.5) {\textcolor{black}{$\circ$}};
\draw[-,dashed] (1.6,1.6) to (1.6,-1.6);

\end{tikzpicture}
\caption{Two wedges left over after modding out the translational symmetry. The $Z_{2}$ symmetry is the reflection with respect to the dashed vertical line.}
\label{pic:orbifold2}
\end{figure}
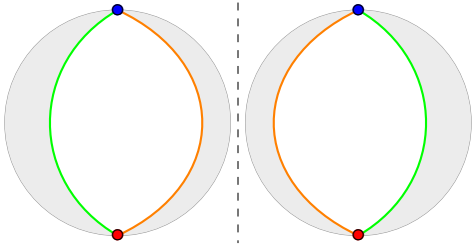

Hence we start with an array of wedges such that adjacent branes of contiguous wedges are of equal tension (see Fig.~\ref{pic:orbifold}). All branes in the array are free to fluctuate with the constraint that the adjacent branes of contiguous wedges fluctuate in the same way. Such an array has a translational symmetry if we jump through two wedges. We can firstly mod out this translational symmetry which leaves us only with two contiguous wedges and the orange branes and green branes are separately fluctuating in the same ways. This is nothing but another way to spell out our situation Equ.~(\ref{eq:action}) in a way that the end-of-the-world (EOW) branes are induced descriptions. Nevertheless, now it is easier for us to orbifold our theory. We have a $Z_{2}$ symmetry that maps the two leftover wedges to one the other (see Fig.~\ref{pic:orbifold2}). If we orbifold our configuration with respect to this symmetry then the orange branes are not allowed to fluctuate (even though it has a nonvanishing tension, similarly to $O$-planes in string theory). As a result, the effective action only contains one dilaton and we have the full effective action
\begin{equation}
    \begin{split}
        S_{\text{orbifold}}&=-\frac{r_{2}-r_{1}}{16\pi G_{3}}\int d^{2}x\sqrt{-g}R[g]-\frac{1}{16\pi G_{3}}\int d^{2}x\sqrt{-g}\delta\phi_{2}(x)\Bigg[R[g]+2\Bigg]\\&-\frac{1}{8\pi G_{3}}\int d^{2}x\sqrt{-g}\Bigg[\frac{T_{2}}{2}\nabla_{\mu}\delta\phi_{2}\nabla^{\mu}\delta\phi_{2}+T_{2}(\delta\phi_{2})^2\Bigg]\,,
    \end{split}
\end{equation}
which reduces to pure JT gravity at the quadratic order in $\delta\phi_{2}$
\begin{equation}
        S_{\text{orbifold}}=-\frac{r_{2}-r_{1}}{16\pi G_{3}}\int d^{2}x\sqrt{-g}R[g]-\frac{1}{16\pi G_{3}}\int d^{2}x\sqrt{-g}\delta\phi_{2}(x)\Bigg[R[g]+2\Bigg]\,,\label{eq:pureJT}
\end{equation}
if we redefine the 2d metric by
\begin{equation}
    g_{\mu\nu}\rightarrow e^{-T_{2}\delta\phi_{2}}g_{\mu\nu}\,.
\end{equation}

However, one may have noticed that if we mod out the translational symmetry in another way that the adjacent branes are green for the two leftover wedges then we would have green branes frozen after orbifolding by the $Z_{2}$ symmetry. Interestingly, it is straightforward to see that if we do this the final action we get is still pure JT gravity although with $\delta\phi_{2}$ replaced by $-\delta\phi_{1}$ (remember $\delta\phi_{1}$ and $\delta\phi_{2}$ are relatively negative). This is the same theory as Equ.~(\ref{eq:pureJT}) if we interpret the dilaton as the fluctuation of the relative position of the two Karch-Randall branes in the ambient space.

\begin{figure}
\centering
\begin{tikzpicture}
\draw[-,black!40] (0,1.5) arc (90:-90:1.5);
\draw[-,black!40] (0,1.5) arc (90:270:1.5);
\draw[-,draw=none,fill=gray!15] (0,1.5) .. controls (-1.2,0.8) and (-1.2,-0.8) .. (0,-1.5) arc (-90:-270:1.5);
\draw[-,thick,green] (0,1.5) .. controls (-1.2,0.8) and (-1.2,-0.8) .. (0,-1.5);

\draw[-,draw=none,fill=gray!15] (0,1.5) .. controls (1.5,0.8) and (1.5,-0.8) .. (0,-1.5) arc (270:450:1.5);
\draw[-,thick,orange] (0,1.5) .. controls (1.5,0.8) and (1.5,-0.8) .. (0,-1.5);
\node at (0,1.5) {\textcolor{blue!100!}{$\bullet$}};
\node at (0,1.5) {\textcolor{black}{$\circ$}};
\node at (0,-1.5) {\textcolor{red!100!}{$\bullet$}};
\node at (0,-1.5) {\textcolor{black}{$\circ$}};

\draw[-,black!40] (-6.4,1.5) arc (90:-90:1.5);
\draw[-,black!40] (-6.4,1.5) arc (90:270:1.5);
\draw[-,draw=none,fill=gray!15] (-6.4,1.5) .. controls (-7.6,0.8) and (-7.6,-0.8) .. (-6.4,-1.5) arc (-90:-270:1.5);
\draw[-,thick,green] (-6.4,1.5) .. controls (-7.6,0.8) and (-7.6,-0.8) .. (-6.4,-1.5);

\draw[-,draw=none,fill=gray!15] (-6.4,1.5) .. controls (-4.9,0.8) and (-4.9,-0.8) .. (-6.4,-1.5) arc (270:450:1.5);
\draw[-,thick,orange] (-6.4,1.5) .. controls (-4.9,0.8) and (-4.9,-0.8) .. (-6.4,-1.5);
\node at (-6.4,1.5) {\textcolor{blue!100!}{$\bullet$}};
\node at (-6.4,1.5) {\textcolor{black}{$\circ$}};
\node at (-6.4,-1.5) {\textcolor{red!100!}{$\bullet$}};
\node at (-6.4,-1.5) {\textcolor{black}{$\circ$}};

\draw[-,black!40] (-3.2,1.5) arc (90:-90:1.5);
\draw[-,black!40] (-3.2,1.5) arc (90:270:1.5);
\draw[-,draw=none,fill=gray!15] (-3.2,1.5) .. controls (-2,0.8) and (-2,-0.8) .. (-3.2,-1.5) arc (-90:90:1.5);
\draw[-,thick,green] (-3.2,1.5) .. controls (-2,0.8) and (-2,-0.8) .. (-3.2,-1.5);
\draw[-,draw=none,fill=gray!15] (-3.2,1.5) .. controls (-4.7,0.8) and (-4.7,-0.8) .. (-3.2,-1.5) arc (-90:-270:1.5);
\draw[-,thick,orange] (-3.2,1.5) .. controls (-4.7,0.8) and (-4.7,-0.8) .. (-3.2,-1.5);
\node at (-3.2,1.5) {\textcolor{blue!100!}{$\bullet$}};
\node at (-3.2,1.5) {\textcolor{black}{$\circ$}};
\node at (-3.2,-1.5) {\textcolor{red!100!}{$\bullet$}};
\node at (-3.2,-1.5) {\textcolor{black}{$\circ$}};

\draw[-,black!40] (3.2,1.5) arc (90:-90:1.5);
\draw[-,black!40] (3.2,1.5) arc (90:270:1.5);
\draw[-,draw=none,fill=gray!15] (3.2,1.5) .. controls (4.4,0.8) and (4.4,-0.8) .. (3.2,-1.5) arc (-90:90:1.5);
\draw[-,thick,green] (3.2,1.5) .. controls (4.4,0.8) and (4.4,-0.8) .. (3.2,-1.5);
\draw[-,draw=none,fill=gray!15] (3.2,1.5) .. controls (1.7,0.8) and (1.7,-0.8) .. (3.2,-1.5) arc (-90:-270:1.5);
\draw[-,thick,orange] (3.2,1.5) .. controls (1.7,0.8) and (1.7,-0.8) .. (3.2,-1.5);
\node at (3.2,1.5) {\textcolor{blue!100!}{$\bullet$}};
\node at (3.2,1.5) {\textcolor{black}{$\circ$}};
\node at (3.2,-1.5) {\textcolor{red!100!}{$\bullet$}};
\node at (3.2,-1.5) {\textcolor{black}{$\circ$}};

\draw[-,dotted,black,thick] (-8.6,0) to (-8.1,0);
\draw[-,dotted,black,thick] (4.9,0) to (5.4,0);
\end{tikzpicture}
\caption{An array of AdS$_3$ wedges such that the adjacent branes for contiguous wedges are of equal tension. In other words, the orange branes and green branes are respectively of equal tension.}
\label{pic:orbifold}
\end{figure}
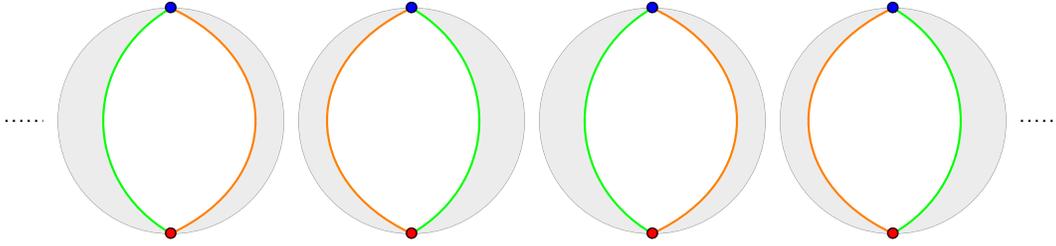

\section{Conformal Symmetry Breaking and the Lifting of the RT Surface Degeneracy }\label{sec:confbreaking}
In the previous section, we showed that the low energy dynamics of the branes are described by  Jackiw-Teitolboim gravity minimally coupled to a scalar field Equ.~(\ref{eq:JT1final}). For the sake of convenience we only consider the pure JT sector from now on. The results can be easily generalized to the general case with the scalar field included.

We know that the AdS$_2$ asymptotic symmetry (the conformal symmetry) prevents the existence of nonzero energy states. Hence to see the emergence of nontrivial physics we have to break the AdS$_2$ asymptotic symmetry. A subgroup of this asymptotic symmetry is the AdS$_2$ isometry the breaking of which is easier to think. However, due to the specific form of the JT gravity where the dilaton field $\phi(x)$ acts as a Lagrange multiplier the geometry is always AdS$_2$ even for the quantum theory.\footnote{However, only when the perturbation theory makes sense i.e. when we still have the 2d geometric picture.} Hence, to break the AdS$_2$ isometry we need a nontrivial profile for the dilaton field. This can be achieved by firstly introducing a spatial cutoff $\partial\mathcal{M}$ of the AdS$_2$ spacetime where we impose the Dirichlet boundary condition for both the dilaton and the metric then sending this cutoff surface to infinity \cite{Maldacena:2016upp}. With this nonzero boundary value of the dilaton $\phi(x)|_{\partial\mathcal{M}}=\phi_{b}$ specified the solution of the equation of motion of the dilaton (coming from the variation with respect to the 2d metric $g_{\mu\nu}$)
\begin{equation}
    \nabla_{\mu}\nabla_{\nu}\phi(x)-g_{\mu\nu}\nabla^{2}\phi+2\phi=0\,,\label{eq:dilatoneom1}
\end{equation}
is not a constant which breaks the AdS$_2$ isometry.


To be precise, the action with appropriate Gibbons-Hawking terms counted turns out to be (see Appendix.~\ref{sec:bdyterms} for an explicit calculation)
\begin{equation}
\begin{split}
 \small   S^{\text{JT}}_{\text{eff}}=&-\frac{\phi_{0}}{16\pi G_{3}}\Big[\int_{\mathcal{M}} d^{2}x\sqrt{-g}R+2\int_{\partial \mathcal{M}}dx\sqrt{-h}K\Big]\\&-\frac{1}{16\pi G_{3}}\Big[\int_{\mathcal{M}} d^{2}x\sqrt{-g}\phi(x)\Big(R+2\Big)+2\phi_{b}\int_{\partial\mathcal{M}}dx\sqrt{-h}K\Big]\,,\label{eq:JT2}
 \end{split}
\end{equation}
where we defined $\phi_{0}=r_{2}-r_{1}$, $K$ is the trace of the extrinsic curvature for the boundary $\partial\mathcal{M}$ and $\phi_{b}$ is the value of the dilaton field $\phi(x)$ on $\partial\mathcal{M}$. 

From the 2d perspective, the cut off we introduced breaks the asymptotic conformal symmetry by introducing nontrivial profiles for the dilaton. Therefore, we expect that the infinite degeneracy of the RT surfaces for the entanglement entropy between the two defects discovered in Sec.~\ref{sec:divergence} will be lifted.\footnote{There we suggested that this degeneracy is related to the  the AdS$_2$ asymptotic symmetry or the conformal symmetry.} In Sec.\ref{sec:2d} we will show that this degeneracy is indeed lifted if we use Equ.~(\ref{eq:JT2}). Interestingly, we will show in Sec.\ref{sec:3d} that from the 3d wedge perspective this lifting has a very nice geometric interpretation due to the brane fluctuation. 

\subsection{The 2d Point of View}\label{sec:2d}
In Sec.~\ref{sec:divergence} we considered the AdS$_3$ wedge with branes fixed so the 2d effective action is just
\begin{equation}
    S_{\text{eff}}=-\frac{\phi_{0}}{16\pi G_{3}}\int_{\mathcal{M}} d^{2}x\sqrt{-g}R\,,
\end{equation}
and so the entanglement entropy functional between the two asymptotic boundaries of the AdS$_2$, or the two defects in the AdS$_3$ wedge, is
\begin{equation}
    S_{\text{EE functional}}=\frac{\phi_{0}}{4G_{3}}\,.\label{eq:func}
\end{equation}
As it is in the standard Ryu-Takayagnagi calculation \cite{Ryu:2006bv}, we have to minimize this functional over a constant time slice of the AdS$_2$. However, we have a constant functional Equ.~(\ref{eq:func}). As a result, any point on a constant time slice of AdS$_2$ contributes is a solution of the minimization.

Naively, this degeneracy is eliminated if we have a time-independent dilaton field which depends on the position along the constant time slice of AdS$_2$ in our effective action Equ.~(\ref{eq:JT2}). In this case, the entanglement entropy is given by
\begin{equation}
    S_{\text{EE}}=\min_{x}\Big(\frac{\phi_{0}+\phi(x)}{4G_{3}}\Big)\,,
\end{equation}
where $x$ denotes the position on a constant time slice of AdS$_2$. 

More precisely, the equations of motion from the action Equ.~(\ref{eq:JT2}) are
\begin{equation}
R+2=0\,,\quad \nabla_{\mu}\nabla_{\nu}\phi=g_{\mu\nu}\phi\,,
\end{equation}
where the second equation is equivalent to Equ.~(\ref{eq:dilatoneom1}). The first equation is solved by the AdS$_2$ global patch
\begin{equation}
    ds^2=d\rho^2-\cosh^{2}(\rho)dt^{2}\,,
\end{equation}
which has two asymptotic boundaries $\rho\rightarrow\pm\infty$.

The second equation is the equation of motion of the dilaton and if we are looking for a time-independent solution it is solved by
\begin{equation}
    \phi(x)=\phi_{r}\sinh(\rho)\,,
\end{equation}
for which however the two asymptotic boundaries $\rho\rightarrow\pm\infty$ are no longer symmetric. Hence we will instead look for time-dependent solutions. Such a solution is given by
\begin{equation}
    \phi(x)=\phi_{r}\cosh\rho\cos t\,,\label{eq:dilatonsol}
\end{equation}
where the two defects ($\rho\rightarrow\pm\infty$) are symmetric.

As a result, to calculate the entanglement entropy between the two defects we have to use the maxmin proposal \cite{Wall:2012uf} for the entropy functional
\begin{equation}
    S_{\text{EE}}=\max_{t} \min_{\rho}\Big(\frac{\phi_{0}+\phi(x)}{4G_{3}}\Big)=\frac{\phi_{0}+\phi_{r}}{4G_{3}}\,,
\end{equation}
which gives a finite entanglement entropy and there is a unique solution $\rho=0,t=0$ for the maximin problem. Therefore the infinite degeneracy is lifted. 

\subsection{The 3d Point of View}\label{sec:3d}
\begin{figure}
\centering
\begin{tikzpicture}
\draw[-,black!40] (0,2) arc (90:-90:2);
\draw[-,black!40] (0,2) arc (90:270:2);

\draw[-,draw=none,fill=gray!15] (0,2) .. controls (-1.5,1) and (-1.5,-1) .. (0,-2) arc (-90:-270:2);
\draw[-,thick,black] (0,2) .. controls (-1.5,1) and (-1.5,-1) .. (0,-2);

\draw[-,draw=none,fill=gray!15] (0,2) .. controls (2,1) and (2,-1) .. (0,-2) arc (270:450:2);
\draw[-,thick,black] (0,2) .. controls (2,1) and (2,-1) .. (0,-2);

\draw[-,thick,green] (0.8,1.5)..controls (0.5,1.2) and (-0.5,1.2)..(-0.7,1.3);
\draw[-,thick,green] (1,1.2)..controls (0.5,0.9) and (-0.5,0.9)..(-0.9,1);
\draw[-,thick,green] (1.2,0.9)..controls (0.5,0.6) and (-0.5,0.6)..(-1,0.7);
\draw[-,thick,green] (1.4,0.6)..controls (0.5,0.3) and (-0.5,0.3)..(-1.1,0.4);
\draw[-,thick,green] (1.5,0) to (-1.1,0);
\draw[-,thick,green] (0.8,-1.5)..controls (0.5,-1.2) and (-0.5,-1.2)..(-0.7,-1.3);
\draw[-,thick,green] (1,-1.2)..controls (0.5,-0.9) and (-0.5,-0.9)..(-0.9,-1);
\draw[-,thick,green] (1.2,-0.9)..controls (0.5,-0.6) and (-0.5,-0.6)..(-1,-0.7);
\draw[-,thick,green] (1.4,-0.6)..controls (0.5,-0.3) and (-0.5,-0.3)..(-1.1,-0.4);
\node at (0,2) {\textcolor{blue!100!}{$\bullet$}};
\node at (0,2) {\textcolor{black}{$\circ$}};
\node at (0,-2) {\textcolor{red!100!}{$\bullet$}};
\node at (0,-2) {\textcolor{black}{$\circ$}};

\end{tikzpicture}
\caption{The empty AdS$_{3}$ wedge consisting of two KR branes. The bulk geometry is the AdS$_3$ global patch. The green curves are representatives of the degenerate RT surfaces calculating the entanglement entropy between the two defects (the blue dot and the red dot). }
\label{pic:emptyads3rig}
\end{figure}
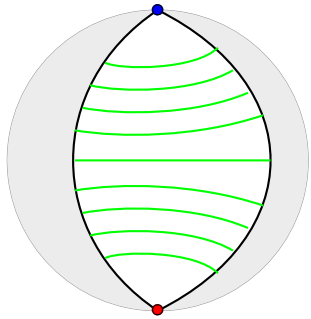

In Sec.~\ref{sec:divergence}, it is noticed that the infinite degeneracy of RT surfaces for the entanglement entropy between the two defects is realized in the AdS$_3$ wedge as the existence of an infinite number of degenerate RT surfaces. These RT surfaces go from one brane to the other, satisfy orthogonal boundary conditions near the brane and they have equal area $A=r_{2}-r_{1}=\phi_{0}$ (see Fig.\ref{pic:emptyads3rig}). This degeneracy of the RT surfaces depends highly on the fact that the shape of the branes are fixed.

\begin{figure}
\centering
\begin{tikzpicture}
\draw[-,black!40] (0,2) arc (90:-90:2);
\draw[-,black!40] (0,2) arc (90:270:2);

\draw[-,draw=none,fill=gray!15] plot [smooth,tension=0.5] coordinates{(0,2)(-0.8,1.2)(-0.8,0.6)(-0.7,0)(-1,-0.9)(0,-2)} (0,-2) arc (-90:-270:2);
\draw[thick] plot [smooth,tension=0.5] coordinates{(0,2)(-0.8,1.2)(-0.8,0.6)(-0.7,0)(-1,-0.9)(0,-2)};
\draw[-,draw=none,fill=gray!15] plot [smooth,tension=0.5] coordinates{(0,2)(1,1.5)(0.8,0.6)(0.7,0)(1.2,-1.1)(0,-2)} (0,-2) arc (270:450:2);
\draw[thick] plot [smooth,tension=0.5] coordinates{(0,2)(1,1.5)(0.8,0.6)(0.7,0)(1.2,-1.1)(0,-2)};

\draw[-,thick,orange] (0.9,1.6)..controls (0.5,1.2) and (-0.5,1.2)..(-0.7,1.3);
\draw[-,thick,orange] (1,1.2)..controls (0.5,0.9) and (-0.5,0.9)..(-0.85,1);
\draw[-,thick,orange] (0.9,0.8)..controls (0.5,0.6) and (-0.5,0.6)..(-0.8,0.7);
\draw[-,thick,orange] (0.8,0.5)..controls (0.5,0.3) and (-0.5,0.3)..(-0.75,0.4);
\draw[-,thick,orange] (0.7,0) to (-0.7,0);
\draw[-,thick,orange] (0.8,-1.5)..controls (0.5,-1.2) and (-0.5,-1.2)..(-0.7,-1.3);
\draw[-,thick,orange] (1.1,-1.2)..controls (0.5,-0.9) and (-0.5,-0.9)..(-0.95,-1);
\draw[-,thick,orange] (1.2,-0.9)..controls (0.5,-0.6) and (-0.5,-0.6)..(-1,-0.7);
\draw[-,thick,orange] (1,-0.5)..controls (0.5,-0.3) and (-0.5,-0.3)..(-0.9,-0.4);
\node at (0,2) {\textcolor{blue!100!}{$\bullet$}};
\node at (0,2) {\textcolor{black}{$\circ$}};
\node at (0,-2) {\textcolor{red!100!}{$\bullet$}};
\node at (0,-2) {\textcolor{black}{$\circ$}};

\end{tikzpicture}
\caption{The shape of the two branes are fluctuating. The orange curves are representatives of the bulk minimal area surfaces. The orange curve are of different length and we pick up the shortest one of them to calculate the entanglement entropy between the two defects (the blue dot and the red dot). }
\label{pic:emptyads3fluc}
\end{figure}
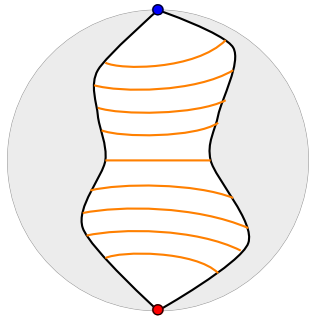

Now we consider fluctuating branes whose shapes are not fixed (see Fig.\ref{pic:emptyads3fluc}). Therefore, the degeneracy of the RT surfaces is automatically broken. The low energy dynamics of the brane fluctuation is controlled by Equ.~(\ref{eq:JT2}). Hence for a time-dependent dilaton profile as in Equ.~(\ref{eq:dilatonsol}) the shapes of the branes are time-dependent. As a result, we have to use the bulk HRT proposal \cite{Hubeny:2007xt} looking for HRT surfaces connecting the two branes. The calculations are expected to be complicated so we will not do it in this paper. But the geometric picture is clear that the degeneracy of the RT surfaces will be lifted.

\section{The Holographic Dual}\label{sec:holographicdual}

In the previous section, we showed that to see the emergence of nontrivial physics we have to take a cutoff in the AdS$_2$, impose Dirichlet boundary conditions and send the cutoff to infinity. From the 3d bulk point of view this is induced by taking a cutoff of the 3d bulk, imposing Dirichlet boundary conditions and sending the cutoff the infinity.

\begin{figure}
\centering
\begin{tikzpicture}
\draw[-,black!40] (0,2) arc (90:-90:2);
\draw[-,black!40] (0,2) arc (90:270:2);
\draw[-,draw=none,fill=gray!15] plot [smooth,tension=0.5] coordinates{(0,2)(-0.8,1.2)(-0.8,0.6)(-0.7,0)(-1,-0.9)(0,-2)} (0,-2) arc (-90:-270:2);
\draw[thick] plot [smooth,tension=0.5] coordinates{(0,2)(-0.8,1.2)(-0.8,0.6)(-0.7,0)(-1,-0.9)(0,-2)};
\draw[-,draw=none,fill=gray!15] plot [smooth,tension=0.5] coordinates{(0,2)(1,1.5)(0.8,0.6)(0.7,0)(1.2,-1.1)(0,-2)} (0,-2) arc (270:450:2);
\draw[thick] plot [smooth,tension=0.5] coordinates{(0,2)(1,1.5)(0.8,0.6)(0.7,0)(1.2,-1.1)(0,-2)};
\draw[-,draw=none,fill=red!15] (-0.684,-1.88) arc (170:10:0.6946)--(-0.684,-1.88) arc (-110:-70:2);
\draw[-,draw=none,fill=red!15] (-0.684,1.88) arc (-170:-10:0.6946)--(-0.684,1.88) arc (110:70:2);
\draw[-,red] (-0.684,1.88) arc (-170:-10:0.6946);
\draw[-,red] (-0.684,-1.88) arc (170:10:0.6946);
\node at (0,2) {\textcolor{blue!100!}{$\bullet$}};
\node at (0,2) {\textcolor{black}{$\circ$}};
\node at (0,-2) {\textcolor{red!100!}{$\bullet$}};
\node at (0,-2) {\textcolor{black}{$\circ$}};
\end{tikzpicture}
\caption{The shape of the two branes are fluctuating. The AdS$_2$ cutoff is induced by the AdS$_3$ cutoff (the red curve). The shaded regions are cutoff. Now the cutoff asymptotic boundary of the wedge are two intervals. }
\label{pic:3dcutoff}
\end{figure}
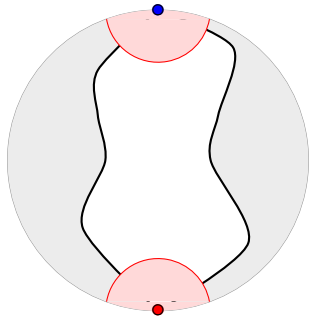

Interestingly, before we send the 3d cutoff to infinity, this 3d cutoff blows up the two asymptotic defects to two intervals (see Fig.~\ref{pic:3dcutoff}). Then sending the 3d cutoff to infinity is equivalent to shrinking these two intervals to zero size. From this observation we can speculate the holographic dual of the dilaton gravity Equ.~(\ref{eq:JT1final}).

 In the configuration described in Fig.~\ref{pic:3dcutoff} the field theory dual lives on the two disconnected intervals introduced by the cutoff below the defects. If we assume the gauge/gravity duality for AdS$_3$/CFT$_2$ where we have Einstein's gravity in the AdS$_3$ bulk then this field theory dual is nothing but a 2d gauge theory with specific boundary conditions on the boundary of the intervals. This is in a sense similar to the AdS/BCFT correspondence \cite{Fujita:2011fp}. Moreover, the gauge theory should be a large-$N$ theory since the bulk is described by Einstein's gravity.
 
 When we send the 3d cutoff to infinity or equivalently shrink the intervals to zero size, we are doing dimensional reduction of the 2d gauge theory and taking the low energy limit. This resulting theory would be the holographic dual of the dilaton gravity Equ.~(\ref{eq:JT1final}) as both of them are the low energy description of the same system- Einstein's gravity in the AdS$_3$ wedge with two fluctuating Karch-Randall branes.
 
 It is interesting to notice that it is a generic feature of gauge theories that when we compactify it on a compact manifold and consider the low energy regime then its finite temperature partition function is captured by a random matrix theory \cite{Aharony:2003sx,Aharony:2005ew} (see \cite{Alvarez-Gaume:2005dvb,Liu:2004vy} for related works). This random matrix is nothing but the Polyakov loop which is the order parameter of the confinement-deconfinement phase transition of the gauge theory and it is the zero mode of the Yang-Mill fields after compactification \cite{Aharony:2003sx}, which therefore describes the low energy dynamics of the system. 
 
 Thus we can conclude with a speculation that the dilaton gravity theory Equ.~(\ref{eq:JT1final}) is dual to such a random matrix theory if we attempt to compute its finite temperature partition function.

\section{A Puzzle About the Energy Spectrum and Its Resolution}\label{sec:spectrumpuzzle}
In Sec.~\ref{sec:spectrumEE} we noticed that when we have fixed branes, we have a conformal quantum mechanics system as the holographic dual for the AdS$_3$ wedge. This quantum mechanics system is not dynamical at all and we only have degenerate ground states with the degeneracy given by the brane positions (the brane positions are uniquely determined by brane tensions by solving the Israel's junction equations \cite{Kraus:1999it,Fujita:2011fp}). In that case, we have the energy density
\begin{equation}
    \rho(E)=e^{\frac{\phi_{0}}{4G_{3}}}\delta(E)\,.\label{eq:energydelta}
\end{equation}

Now with the conformal symmetry breaking and the nontrivial dynamics introduced in Equ.~(\ref{eq:JT2}), we can ask for the new energy spectrum which is the low energy spectrum of the branes dynamics in the AdS$_3$ wedge (in this section we focus on the set up discussed in Sec.~\ref{sec:orbifolding}). With the boundary value of the dilaton field $\phi_{b}$ fixed, the dynamics of Equ.~(\ref{eq:JT2}) happens purely on the boundary $\partial\mathcal{M}$. This is simply because the path integral of the dilaton field $\phi(x)$ imposes the constraint
\begin{equation}
    R+2=0\,,\label{eq:eomspacetime}
\end{equation}
and the first term in the action Equ.~(\ref{eq:JT2}) is given by the Euler number $\chi(\mathcal{M})$ of the manifold $\mathcal{M}$. To get the energy spectrum, we have to firstly compute the Euclidean partition of the theory. In the Euclidean signature, the action can then be written as
\begin{equation}
    S_{\text{eff}}=-\frac{\phi_{0}}{4G_{3}}\chi(\mathcal{M})-\frac{\phi_{b}}{8\pi G_{3}}\int_{\partial\mathcal{M}}dx\sqrt{h}K\,.\label{eq:Seff}
\end{equation}
The 2d spacetime manifold $\mathcal{M}$ satisfies Equ.~(\ref{eq:eomspacetime}) and to compute the energy spectrum we take its geometry to be Euclidean AdS$_2$
\begin{equation}
    ds^2=(\frac{2\pi}{\beta})^2\Big[d\rho^2+\sinh^{2}(\frac{2\pi\rho}{\beta})d\tau^{2}\Big]\,,
\end{equation}
where the Euclidean time $\tau$ is periodic with period $\beta$. In this background the solution of the dilaton equation of motion is given by
\begin{equation}
    \phi(x)=\frac{2\pi \phi_{r}}{\beta}\cosh(\frac{2\pi\rho}{\beta})\,,
\end{equation}
which is zero at zero temperature. 

Moreover, with the cutoff given by the closed curve $(\rho(u),\tau(u))$, the topology of $\mathcal{M}$ is a disk with Euler number
\begin{equation}
    \chi(\mathcal{M})=1\,.
\end{equation}
We fix the induced metric on the boundary curve $(\rho(u),\tau(u))$ and the value of the dilaton field $\phi_{b}$ on it using a constant $\phi_{r}$ as
\begin{equation}
    ds_{\text{bdy}}^2=\frac{4\pi^2}{\beta^2}
    \Big[\rho'(u)^2+\sinh^2(\frac{2\pi\rho}{\beta})\tau'(u)^2\Big]du^2=\frac{du^2}{\epsilon^2}\,,\quad\phi_{b}=\frac{\phi_{r}}{\epsilon}\,,
\end{equation}
where $u$ goes from $0$ to $\beta$ and $\epsilon$ is a UV cutoff length scale. Therefore we have the trace of the extrinsic curvature of the boundary curve
\begin{equation}
    \begin{split}\label{eq:Kepsilon}
        K&=\frac{\beta}{2\pi}\frac{\rho'\tau''-\tau'\rho''+\frac{2\pi}{\beta}\tau'^{3}\sinh(\frac{2\pi\rho}{\beta})\cosh(\frac{2\pi\rho}{\beta})+\frac{4\pi}{\beta}\tau'\rho'^{2}\coth(\frac{2\pi\rho}{\beta})}{(\rho'^2+\frac{4\pi^2}{\beta^2}\sinh^{2}(\frac{2\pi\rho}{\beta})\tau'^2)^{\frac{3}{2}}}\sinh(\frac{2\pi\rho}{\beta})\,\\
        &=1+\Big[\frac{2\pi^2\tau'^2}{\beta^2}+\frac{\tau'''}{\tau'}-\frac{3\tau''^2}{2\tau'^2}\Big]\epsilon^2+\mathcal{O}(\epsilon^3)\,,
    \end{split}
\end{equation}
where it is useful to notice that to our order of approximation we have $\rho=\frac{\beta}{2\pi}\ln(\frac{\beta}{\pi\epsilon\tau'})$. As a result, the effective action can be written as
\begin{equation}
\begin{split}
    S_{\text{eff}}&=-\frac{\phi_{0}}{4G_{3}}-\frac{\phi_{r}}{8\pi G_{3}}\int_{0}^{\beta}du\Big(\frac{2\pi^2\tau'^2}{\beta^2}+\frac{\tau'''}{\tau'}-\frac{3\tau''^2}{2\tau'^2}\Big)\,\\
    &=-\frac{\phi_{0}}{4G_{3}}-\frac{\phi_{r}}{8\pi G_{3}}\int_{0}^{\beta}du\quad\text{Sch}(\tan\frac{\pi}{\beta}\tau(u),u)\,,\label{eq:Schwarzian}
    \end{split}
\end{equation}
where we sent the cutoff the infinity (i.e. $\epsilon\rightarrow0$) and neglected higher order terms in $\epsilon$, we dropped a constant term $\propto\frac{1}{\epsilon^2}$ and $\text{Sch}$ denotes the Schwarzian \cite{Maldacena:2016hyu,Maldacena:2016upp}. As it is proved in \cite{Stanford:2017thb}, the partition function of the Schwarzian theory is one-loop exact. As a consequence, we can compute the energy spectrum by computing the partition function and then Laplace transform the result \cite{Saad:2019lba,Stanford:2017thb} getting
\begin{equation}
    \rho(E)\propto e^{\frac{\phi_{0}}{4G_{3}}}\sinh(2\pi\sqrt{\frac{\phi_{r}}{4G_{3}}E})\,.
\end{equation}
We emphasize this is the exact energy spectrum of the Schwarzian theory Equ.(\ref{eq:Schwarzian}) \cite{Stanford:2017thb} and it describes the perturbative (with no nontrivial topology for the AdS$_2$ manifold) low energy spectrum of the nearly AdS$_2$ (NAdS$_2$) description Equ.~(\ref{eq:JT2}).

However, this formula is very puzzling because if we take low energy limit of the brane dynamics, i.e. $\phi_{r}\rightarrow0$ or $E\rightarrow0$, we wouldn't recover the energy spectrum Equ.~(\ref{eq:energydelta}) and we instead get zero. This is not consistent with the standard Wilsonian picture.

Before we try to resolve this puzzle, let's make sure that this puzzle is itself well-defined. This puzzle is based on the observation that sending the boundary profile of the dilaton $\phi_r$ (remember $\phi_{b}=\frac{\phi_{r}}{\epsilon}$) to zero the Karch-Randall branes should be frozen and we recover the conformally symmetric set up with two rigid branes. This is because in the effective JT description, the dilaton equation of motion Equ.~(\ref{eq:dilatoneom1}) has only vanishing solutions if the dialton is constrained to vanish near the asymptotic boundary. However, one may have noticed that we get the effective JT description because we only consider small brane fluctuations. Interestingly, as we send the cutoff scale $\epsilon$ to 0, it is not obvious that the brane fluctuation is always small as close to the cutoff slice $\phi\sim\phi_{b}=\frac{\phi_{r}}{\epsilon}$. Hence we have to carefully restore the dependence of the AdS$_3$ length scale $l_{AdS}$. With this parameter restored, small brane fluctuation means that $\frac{\phi}{\l_{AdS}}\ll1$ or more stringently $\frac{\phi_{b}}{l_{AdS}}=\frac{\phi_{r}}{\epsilon l_{AdS}}\ll1$. Thus, our derivation of the effective JT description works in the limit that we always take $\frac{\phi_{r}}{l_{AdS}}\rightarrow0$ before we take $\epsilon\rightarrow0$. Or more precisely, we always take $\frac{\phi_{r}}{l_{AdS}}\sim O(\epsilon^2)$. With this limit carefully refined, our derivation of the effective JT gravity description is valid and we can consistently send the brane fluctuation to zero by setting $\frac{\phi_{r}}{l_{AdS}}$ to zero first. In other words, using our bulk AdS$_3$ picture, we provide a UV-completion of JT gravity as a trivial conformal quantum mechanics system and JT gravity is an irrelevant deformation of this conformal quantum mechanics system. From the AdS$_{3}$ bulk perspective, this is consistent with the proposal of the holographic dual of the $T\bar{T}$ deformation \cite{McGough:2016lol}, which is an irrelevant deformation of CFT$_{2}$ with the holographic dual as a UV-cutoff AdS$_{3}$. We also emphasize that this wouldn't destroy the reduction from JT gravity to the Schwarzian description Equ.~(\ref{eq:Schwarzian}). The reason is that the whole boundary action $-\frac{\phi_{b}}{8\pi G}\int d^{2}x\sqrt{-h}K$ (see Equ.~(\ref{eq:Seff})) is proportional to $\phi_{b}$ and therefore the $\epsilon$-counting in the expansion of the boundary extrinsic curvature $K$ in Equ.~(\ref{eq:Kepsilon}) is not affected even when $\phi_b$ scales with $\epsilon$.

To resolve this puzzle, we have to consider the finite temperature partition function,the Laplacian transform of which gives us the energy 
spectrum. To the lowest order in the topological expansion, we only have to consider the Schwarzian theory Equ.~(\ref{eq:Schwarzian}). The resulting partition function is one-loop exact and it is obtained in \cite{Stanford:2017thb} as
\begin{equation}
    Z(\beta)=\frac{1}{\beta^{\frac{3}{2}}}e^{\frac{\pi\phi_{r}}{4\beta G_{3}}+\frac{\phi_{0}}{4G_{3}}}\,,
\end{equation}
where $\beta$ is the inverse temperature. For convenience let's consider $\ln Z(\beta)$ instead
\begin{equation}
    \ln Z(\beta)=\frac{\pi\phi_{r}}{4\beta G_{3}}+\frac{\phi_{0}}{4G_{3}}-\frac{3}{2}\ln\beta\,.
\end{equation}
From here we can see that if we take the $G_{3}\rightarrow0$ limit for the bulk Einstein's gravity then we will indeed get the $\ln Z(\beta)=\frac{\phi_{0}}{4G_{3}}$ for the conformal quantum mechanics Equ.~(\ref{eq:energydelta}) in the limit that $\phi_{r}\rightarrow0$ as long as $\beta$ is not nonperturbatively large ($\sim\mathcal{O}(e^{\frac{\phi_{0}}{4G_{3}}})$) or small ($\sim\mathcal{O}(e^{-\frac{\phi_{0}}{4G}})$). This is because in these two limits the 2d geometric picture breaks down as the genus expansion of the JT gravity diverges at large genus number $g$. The reason is that in the first case the growing of the Weil-Peterson volume in $g$ will eventually destroy the perturbative expansion in $g$ (see \cite{Saad:2019lba} for explicit formulas of the Weil-Peterson volume and the perturbative expansion in $g$)\footnote{We thank Kristan Jensen for explaining this to us.} and in the second case the spacetime degenerates below the Planck size. Hence in these two limits we cannot ignore higher genus effects and we may also have to include other nonperturbative objects like defects \cite{Mertens:2019tcm,Maxfield:2020ale} and branes \cite{Blommaert:2019wfy,Okuyama:2021eju,Blommaert:2021fob,Blommaert:2021gha}. Hence we can see that in the appropriate regime where our theory is well-defined we indeed recover the conformal quantum mechanics result of the energy spectrum Equ.~(\ref{eq:energydelta}) in the limit $\phi_{r}\rightarrow0$ i.e. when we turn off the brane fluctuation.

We would like to summarize our point as that, without supersymmetry, Schwarzian quantum mechanics is not enough to capture the whole low energy physics of the bulk gravitational theory. We find that the low energy topological sector (the delta function in the energy spectrum) doesn't manifest in the exact Schwarzian energy spectrum. Although with supersymmetry it is known that the super-Schwarzian quantum mechanics can capture this topological sector \cite{Stanford:2017thb}, it is not clear whether the whole topological sector is captured. The topological sector should manifest itself in the long time behavior of the system.\footnote{See \cite{Lin:2022rzw,Lin:2022zxd} for a recent exploration of this idea in the supersymmetric case.}

\section{Discussions and Conclusions}\label{sec:conclusion}
In this paper, we studied quantum aspects of two types of quantum gravity theories in (nearly) AdS$_2$ by embedding them into the Karch-Randall braneworld. The basic setup we considered is the same as the recently proposed wedge holography \cite{Akal:2020wfl,Miao:2020oey} where we have two Karch-Randall branes forming a wedge. In our case, the bulk geometry is asymptotically AdS$_3$ so the field theory dual is supported by two disconnected defects on the asymptotic boundary of the bulk where the two branes intersect.

The first theory we studied is the 2d Einstein-Hilbert gravity. This theory is purely topological and we found that it can be realized in the standard wedge holography setup, where we have two rigid Karch-Randall branes, as the two dimensional effective description of the bulk Einstein's gravity. Moreover, using the three equivalent descriptions of the Karch-Randall braneworld (as we reviewed in the introduction) we can deduce that this theory duals to a conformal quantum mechanics system supported on the two disconnected defects. We studied the entanglement entropy between these two defects using Ryu-Takayanagi formula in both the 3d description and the 2d description and we found that their results precisely match. Interestingly, we noticed that this entanglement entropy is the same for both the ground state and the thermofield double state which tells us that the defect system only has degenerate ground states and from here with the fact that the system is conformally invariant we can get the precise energy spectrum of the defect system. This is consistent with the fact that 2d Einstein-Hilbert gravity is pure topological with no interesting dynamics. We further confirmed this expectation by studying the L/R entropy introduced in \cite{Geng:2020fxl} whose time dependence in general captures interesting dynamical properties of the system. We found that the L/R entropy is totally time-independent. Furthermore, we observed that there is an infinite degeneracy of the RT surfaces for the entanglement entropy between the two defects and we suggested that this degeneracy is intimately related to the fact that the system is conformally invariant.

The second theory we studied the a type of 2d dilaton gravity. More precisely, it is the Jackiw-Teitelboim gravity minimally coupled with a massive scalar field. We found that the pure Jackiw-Teitelboim gravity emerges in two specific limits where we take the two branes to be tensionless or we carefully orbifold our setup. This class of theories is realized by allowing the two Karch-Randall branes to fluctuate and they describe the low energy dynamics of the branes' fluctuation.\footnote{See \cite{Suzuki:2022yru,Izumi:2022opi} for recent studies of other aspects of the brane dynamics.} Such theories generically breaks the conformal symmetry or the AdS$_2$ isometry due to the nontrivial dilaton profile. We found that in this case the infinite degeneracy of RT surfaces for the entanglement entropy between the two defects is lifted. This confirms our previous suggestion that this degneracy is intimately related to the conformal invariance. Finally, we speculated that this class of dilaton theories duals to random matrix theories if we assume gauge/gravity duality in the 3d bulk.

At the end, we resolved a puzzle regarding the energy spectrum of the dilaton gravity theories the 2d Einstein-Hilbert gravity. The later should be the low energy limit of the former based on our picture that besides the Einstein's gravity in the AdS$_3$ wedge the dilaton gravity theories captures the low energy dynamics of the branes. We found that the thermal partition of the former reduces to that of the later in the appropriate limit taking $\phi_{r}\rightarrow0$ in the regime when the 2d geometric picture is valid. 

\section*{Acknowledgments}
We are grateful to Andreas Blommaert, Andreas Karch, Kristan Jensen, Hong Liu, Suvrat Raju, Lisa Randall, Subir Sachdev and Douglas Stanford for useful discussions. We would like to thank Andreas Blommaert, Andreas Karch, Hong Liu, Suvrat Raju and Lisa Randall for comments on the draft. We thank Andreas Karch, Carlos Perez-Pardavila, Suvrat Raju, Lisa Randall, Marcos Riojas and Sanjit Shashi for relevant collaborations. HG is very grateful to his parents and recommenders. This work is supported by the grant (272268) from the Moore Foundation ``Fundamental Physics from Astronomy and Cosmology."

\begin{appendices}

\section{Boundary Terms of the 2d Dilaton Theory}\label{sec:bdyterms}
In this section, we derive the boundary terms for the 2d dilaton gravity theory from the 3d Gibbons-Hawking term. As we discussed in Sec.~\ref{sec:holographicdual}, the cutoff of the 2d geometry is induced from a cutoff of the 3d bulk (see Fig.~\ref{pic:3dcutoff}) with Dirichlet boundary condition imposed. More precisely, the 3d cutoff is specified by the following boundary conditions for the bulk metric and brane fluctuations
\begin{equation}
    \begin{split}
        ds^{2}|_{\text{bdy}}&=dr^{2}-\cosh^{2}(r)\frac{du^{2}}{\epsilon^{2}}\,,\\
        \delta \phi_{1}(x)|_{\text{bdy}}&=\phi_{b1}=\frac{\phi_{r1}}{\epsilon}\,,\quad \delta \phi_{2}(x)|_{\text{bdy}}=\phi_{b2}=\frac{\phi_{r2}}{\epsilon}\,,\label{eq:dirichlet}
    \end{split}
\end{equation}
where $u$ is the proper time of the intersection between this cutoff boundary and the branes, $\epsilon\rightarrow0$ denotes the cutoff scale and $\phi_{b1}$ and $\phi_{b2}$ are constant. The bulk Gibbons-Hawking term is
\begin{equation}
S_{\text{GH}}=-\frac{1}{8\pi G_{3}}\int d^{2}x_{\text{bdy}}\sqrt{-h^{\text{bdy}}}K^{(3)}\,,
\end{equation}
where $h^{\text{bdy}}_{ab}$ is the induced metric on this cutoff boundary which is given by the first equation of Equ.~(\ref{eq:dirichlet}) and $K^{(3)}$ is the trace of the extrinsic curvature of this cutoff boundary in the 3d bulk. It is easy to show that
\begin{equation}
    K^{(3)}=\frac{1}{\cosh^2(r)}K\,,
\end{equation}
where $K$ is the trace of the extrinsic curvature of the boundary in the 2d geometry that appears in Equ.~(\ref{eq:JT2}). This equation is simply due to the fact that the unit normal vector of the cutoff boundary has no $r$-component. Hence we can see that the Gibbons-Hawking terms can be written as
\begin{equation}
    \begin{split}
        S_{\text{GH}}&=-\frac{1}{8\pi G_{3}}\int_{r_{1}+\delta \phi_{1}|_{\text{bdy}}}^{r_{2}+\delta\phi_{2}|_{\text{bdy}}} dr\int \frac{du}{\epsilon}\cosh(r)\frac{K}{\cosh(r)}\,,\\&=-\frac{r_{2}-r_{1}}{8\pi G_{3}}\int \frac{du}{\epsilon}K-\frac{1}{8\pi G_{3}}\int\frac{du}{\epsilon} (\delta\phi_{2}-\delta\phi_{1})|_{bdy}K\,,\\&=-\frac{2\phi_{0}}{16\pi G_{3}}\int_{\partial\mathcal{M}} dx\sqrt{-h}K-\frac{2\phi_{b}}{16\pi G_{3}}\int_{\partial\mathcal{M}}dx\sqrt{-h}K\,,
    \end{split}
\end{equation}
which are precisely the boundary terms appearing in Equ.~(\ref{eq:JT2}). 

\hfill\begin{minipage}{1\linewidth}   \quad\quad We should emphasize that in our case we might have to be careful about the gravitational corner terms (or Hayward terms \cite{Akal:2020wfl}) in the action associated with $\partial\mathcal{M}$---the intersection between the branes and the 3d cutoff boundaries. However, a careful analysis shows that this term doesn't contribute at order $O(\epsilon^{0})$ for the effective action of the boundary soft mode $\tau(u)$. Hence the Schwarzian action Equ.~(\ref{eq:Schwarzian}) is not affected.
\end{minipage}
\end{appendices}
\bibliographystyle{JHEP}
\bibliography{main}
\end{document}